\documentclass[fleqn,usenatbib]{mnras}
\usepackage{mathptmx}

\usepackage[T1]{fontenc}

\usepackage{graphicx}	
\usepackage{amsmath}	
\usepackage{amssymb}	

\usepackage[dvipsnames]{xcolor}
\usepackage{soul}
\usepackage{caption}
\usepackage{footnote}

\newcommand{\Teff}{\mbox{$T_{\mathrm{eff}}$}}
\newcommand{\logg}{\mbox{$\log g$}}

\newcommand{\Line}[3]{#1\,{\sc #2}~$\lambda$#3}

\newcommand{\Ion}[2]{#1\,{\sc #2}}

\newcommand{\Rwd}{\mbox{$R_{\mathrm{WD}}$}}
\newcommand{\Mwd}{\mbox{$M_{\mathrm{WD}}$}}

\newcommand{\Msun}{\mbox{$\mathrm{M}_{\odot}$}}
\newcommand{\Rsun}{\mbox{$\mathrm{R}_{\odot}$}}
\newcounter{tref}

\newcommand{\target}{GD\,424}
\newcommand{\htohe}{\mbox{$\log(\mathrm{H/He})$}}

\title[\target\ -- a highly metal-polluted white dwarf]{GD\,424~--~a helium-atmosphere white dwarf with a large amount of trace hydrogen in the process of digesting a rocky planetesimal}

\author[P. Izquierdo et al.]{Paula Izquierdo$^{1,2}$\thanks{E-mail: pizdo@iac.es},
Odette Toloza$^{3}$,
 Boris T. G\"ansicke$^{3,4}$,
 Pablo Rodr\'iguez-Gil$^{1,2}$,
 \newauthor
 Jay Farihi$^{5}$,
 Detlev Koester$^{6}$,
 Jincheng Guo$^{5}$ and 
 Seth Redfield$^{7}$
\\
$^{1}$Instituto de Astrof\'isica de Canarias, 38205 La Laguna, Tenerife, Spain\\
$^{2}$Departamento de Astrof\'isica, Universidad de La Laguna, 38206 La Laguna, Tenerife, Spain\\
$^{3}$Department of Physics, University of Warwick, Coventry CV4 7AL, UK\\
$^{4}$Center for Exoplanets and Habitability, University of Warwick, Coventry CV4 7AL, UK\\
$^{5}$Physics and Astronomy, University College London, London, WC1E 6BT, UK\\
$^{6}$Institut f\"ur Theoretische Physik und Astrophysik, Universit\"at Kiel, 24098, Kiel, Germany\\
$^{7}$Department of Astronomy and Van Vleck Observatory, Wesleyan University, Middletown, CT 06459, USA
}

\date{Accepted 21 December 2020}

\pubyear{2020}

\begin{document}
\label{firstpage}
\pagerange{\pageref{firstpage}--\pageref{lastpage}}
\maketitle

\begin{abstract}
The photospheric metal pollution of white dwarfs is now well-established as the signature of the accretion of planetary debris. However, the origin of the trace hydrogen detected in many white dwarfs with helium atmospheres is still debated. Here, we report the analysis of \target: a metal-polluted, helium-atmosphere white dwarf with a large amount of trace hydrogen. We determined the atmospheric parameters using a hybrid analysis that combines the sensitivity of spectroscopy to the atmospheric composition, \htohe, with that of photometry and astrometry to the effective temperature, \Teff, and surface gravity, \logg. The resulting white dwarf mass, radius, and cooling age are $\Mwd=0.77\pm0.01\,\Msun$, $\Rwd=0.0109\pm0.0001\,\Rsun$, and $\tau_\mathrm{cool}=215\pm10$~Myr, respectively. We identified and measured the abundances of 11 photospheric metals and argue that the accretion event is most likely either in the increasing or steady state, and that the disrupted planetesimal resembles either CI chondrites or the bulk Earth in terms of its composition. We suggest that the observed $1.33\times 10^{22}$\,g of trace hydrogen in \target\ were at least partly acquired through accretion of water-rich planetary debris in an earlier accretion episode.

\end{abstract}






\section{Introduction}

The high surface gravity of white dwarfs results in elements heavier than helium settling out of their photospheres on time-scales that are much shorter than their cooling ages \citep{schatzman48}. Consequently, their atmospheres are mainly composed of hydrogen or helium\footnote{Hence, the spectra of white dwarfs with hydrogen- and helium-rich atmospheres hotter than $\simeq5000$\,K and $\simeq10\,000$\,K, respectively, are dominated by Balmer (DA white dwarfs) or helium (DB white dwarfs) lines. However, an important caveat is that because of the strong opacity of hydrogen, even some helium-dominated white dwarfs appear as DAs \citep[e.g.][]{kawka05,koesteretal05-1}.}.
However, between 25 and 50 per cent of all white dwarfs show traces of metals in their spectra \citep{Zuckerman03,zuckerman10-1,Koester14}, and it is now firmly established that the origin of these metals is the accretion of tidally disrupted planetary bodies \citep{jura03, verasetal14-1}. As a result, the spectroscopic analysis of metal-polluted white dwarfs has emerged as a powerful tool to measure the bulk compositions of the parent bodies, most of which broadly resemble inner Solar System objects \citep[see e.g.][]{zuckerman07,koester09-1,klein10,gaensicke12,farihietal13-2,raddietal15-1,hollands17,hollands18-2}.

While photospheric metal pollution is unambiguously linked to the accretion of planetary debris, the origin of the trace hydrogen detected in up to 75~per cent of DBs \citep{koester+kepler15-1} is still under debate. As DBs cool, they rapidly develop deep outer convection zones (CVZ, e.g. fig.\,3 in \citealt{bergeronetal11-1}). Material accreted into the photosphere is quickly and homogeneously mixed throughout the CVZ. Metals will diffuse out of the CVZ on time-scales that depend on the physical conditions at its bottom, which rapidly increase with cooling age, from $\tau_\mathrm{diff}=100-1000$\,yr for $\tau_\mathrm{cool}\simeq10$\,Myr to $\tau_\mathrm{diff}\simeq1$\,Myr for $\tau_\mathrm{cool}\simeq500$\,Myr  (\citealt{koester09-1}, see fig.\,1 in \citealt{wyattetal14-1}). In contrast, hydrogen never diffuses out of the CVZ and, consequently, accretion of this element will result in a long-term increase of its total mass, $M_\mathrm{H}$, held in the CVZ. Several possible explanations have been put forward, including primordial hydrogen left over from the evolution of the white dwarf progenitor, accretion from an external source (either the interstellar medium, \citealt{alcock+illarionov80-1}, or planetary bodies, \citealt{veras14}), and most recently, dredge-up of hydrogen from deep envelope layers \citep{rollandetal20-1}.

In order to explore the possible link between the presence of trace hydrogen in DBs and the accretion of planetary debris, \citet{gentile17} used a sample of 729 helium-dominated white dwarfs provided by the Sloan Digital Sky Survey (SDSS) and found that the presence of hydrogen is almost twice more common in metal-polluted DBs than in \textit{pure} DBs. This result suggests that a fraction of the hydrogen may have been acquired through accretion of planetesimals, analogous to the trace metals detected in many white dwarfs. An independent evidence for planetary bodies at least contributing to the trace hydrogen in DBs is the detection of an oxygen excess in a small number of metal-polluted white dwarfs, which has been interpreted as accretion of water (and hence hydrogen) rich material \citep{farihietal13-2,raddietal15-1}.


In this paper we present the analysis of \target, a newly identified metal-polluted DB white dwarf with a large amount of trace hydrogen.
Section~\ref{sec:obs_data} describes the spectra obtained and the data reduction process. In Section~\ref{sec:atm_params} we explain the methodology used to derive the photospheric parameters and chemical composition of \target. We discuss the current state of the accretion episode, the properties of the parent body and the link with trace hydrogen in DBs in Section~\ref{sec:discussion}. Finally, the conclusions are presented in Section~\ref{sec:conclusions}.

\section{Observations and data reduction} 
\label{sec:obs_data}

\subsection{William Herschel Telescope spectra}
We observed \target\ on 2017 August 26 with the Intermediate-dispersion Spectrograph and Imaging System (ISIS) mounted on the Cassegrain focus of the 4.2-m William Herschel Telescope (WHT) located at the Observatorio del Roque de los Muchachos on La Palma, Spain. The ISIS spectrograph is capable of obtaining simultaneous blue and red optical spectra. The blue and red arms are equipped with the EEV12 and the deep-depletion RED+ 2048\,$\times$\,4096\,pixel CCD detectors, respectively. We binned both detectors by two in the spatial and spectral directions. To separate the blue and red spectra we employed the standard 5300-\AA\ dichroic. We used a slit width of 1\,arcsec at the parallactic angle and the R600B and R600R gratings for the blue and red arms, respectively, with two different central wavelengths per grating: 3930 and 4540~\AA\ in the blue, and 6562 and 8200~\AA\ in the red. We also placed the GG495 second-order sorter filter in the red arm. These set-ups provided spectral resolutions (full-width at half-maximum, FWHM) of 1.9~\AA\ for the two blue central wavelengths and 1.7~\AA\ for the red ones. We took two 900-s spectra for each central wavelength. The observations were conducted under clear sky conditions with variable seeing between 0.7 and 1.0 arcsec.

The spectra were bias and flat-field corrected making use of standard procedures within {\sc iraf}\footnote{{\sc iraf} is distributed by the National Optical Astronomy Observatories (NOAO).}, and the cosmic rays removed with the {\sc iraf} package L.A.Cosmic \citep{vandokkum01}. To obtain the 1D spectra we subtracted the sky background and then performed an optimal extraction \citep{horne86-1} with the {\sc starlink}/{\sc pamela}\footnotemark\ data reduction software \citep{marsh89}. The pixel-wavelength solution was obtained by fitting CuNe+CuAr arc lamp spectra with fourth-order polynomials. We also took spectra of the spectrophotometric standard star BD+33\,2642 to correct for the spectral response. These last two steps were performed using {\sc molly}\footnotemark[\value{footnote}].  

\begin{figure*}
\centering
	\includegraphics[width=2.1\columnwidth]{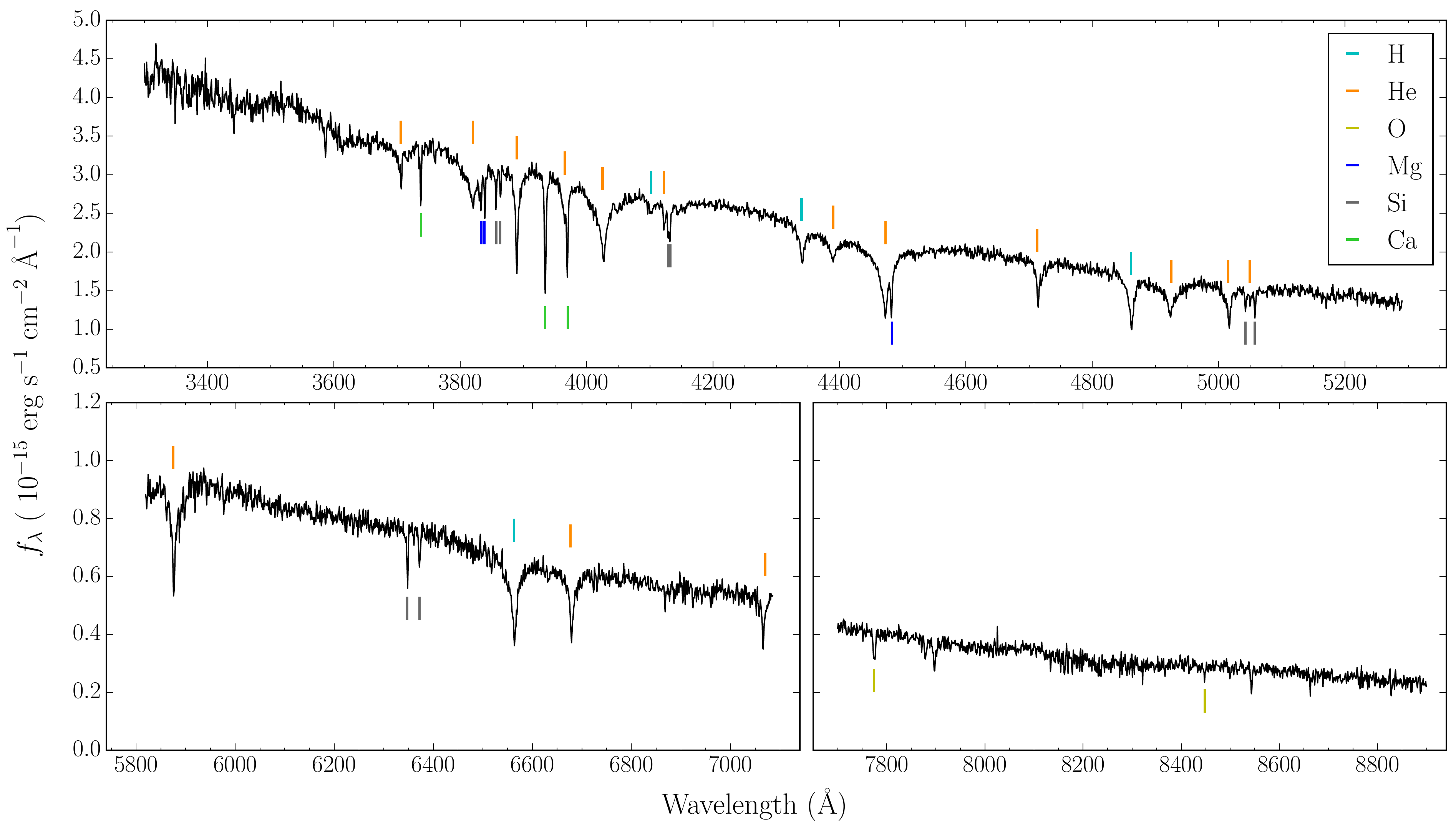}
    \caption{Average spectrum of the DBAZ white dwarf \target\ obtained with WHT/ISIS on 2017 August 26. The spectrum shows broad and strong absorption lines of He and H, and narrower absorption features of O, Mg, Si and Ca.} 
    \label{fig:spec1}
\end{figure*}

We show the composite average spectrum of \target\ in Fig.\,\ref{fig:spec1}. The spectrum shows a helium-dominated photosphere with presence of hydrogen and a number of much narrower metallic absorption lines of oxygen, magnesium, silicon, and calcium, indicating that \target\ is a DBAZ white dwarf (the Z accounts for the presence of metals).

\footnotetext{Both the {\sc pamela} and {\sc molly} packages were developed by Tom Marsh; \url{http://deneb.astro.warwick.ac.uk/phsaap/software/molly/html/INDEX.html}}

\subsection{Keck spectra}

A high-resolution spectrum ($\mathrm{FWHM}=0.1$\,\AA) of the target was obtained on 2019 January 9 using the High Resolution Echelle Spectrometer \cite[HIRES,][]{vogt94} on the 10-m Keck\,I telescope at Mauna Kea Observatory, Hawaii. HIRES is a cross-dispersed \'echelle spectrograph equipped with a mosaic of three MIT--LL $2048 \times 4096$\, pixel CCD detectors. To carry out the observation we used the HIRESb configuration with the C5 decker ($1.148 \times 7$\,arcsec slit), covering the spectral range $3100-5950$\,\AA\ at a nominal spectral resolving power of $R\approx37\,000$. The data were acquired in good conditions at airmasses between 1.7 and 1.8 and were taken in three consecutive exposures of 1800\,s each. The spectra were initially reduced using standard tasks within {\sc iraf} to create master bias and flat frames. Extraction and further reduction was performed using {\sc makee}\footnote{\url{http://www.astro.caltech.edu/~tb/makee/}}. The night in question lacked a bright trace star for the extraction, and thus observations of BD+28\,4211 taken on 2016 June 3 with the same instrument settings were used for this purpose. Each of the three exposures were reduced separately, as were the data from the three individual instrument arrays. Standard procedures were followed in {\sc makee}, including bias subtraction, flat-fielding, order definition, extraction, sky subtraction, wavelength calibration, and heliocentric velocity corrections.

The Keck spectra display broad helium and hydrogen absorption lines (average FWHM of $\simeq 10$\,\AA) and many narrow metal lines (average FWHM of $\simeq 0.2$\,\AA) also in absorption. The much higher spectral resolution and bluer coverage provided by HIRES allowed us to identify additional metal transitions produced by sodium, aluminium, chromium, manganese, iron and nickel, as well as magnesium, silicon and calcium that were also detected in the WHT spectrum.

\section{Atmospheric parameters}
\label{sec:atm_params}

Measurements of the effective temperatures (\Teff) and surface gravities (\logg) of DB white dwarfs can be obtained from either spectroscopy or photometry. Making use in both cases of synthetic atmospheric spectra, the spectroscopic method relies on fitting the helium and Balmer lines \citep[e.g.][]{vossetal07-1, bergeronetal11-1}, while the photometric method is based on reproducing the observed photometry from the synthetic model spectra \citep{gentile19}, the observed parallax of the sources, and the well-established white dwarf mass-radius relation. However, these methods do usually not arrive at the same solution: \Teff\ derived from the photometry alone is found systematically lower, while the \logg\ values are different but do not show a general trend \citep{genest-beaulieu+bergeron19-1,tremblayetal19-2}. Here, we will explore the results of both methods applied to \target\ and, in addition, implement a hybrid method that simultaneously makes use of the independent constraints provided by both the photometric and spectroscopic data.


However, an additional complication in the case of \target\ are the photospheric trace abundances of both hydrogen and several metals, which do affect the structure of the photosphere in terms of providing free electrons and additional opacities \citep[e.g.][]{hollands17}. For the effective temperature of \target\  (16\,560\,K, see Section~\ref{sec:hybridapp}), the effect on the ionisation balance is relatively minor. However, line blanketing, in particular of the many strong metal lines in the ultraviolet\footnote{For examples of ultraviolet spectra of DB white dwarfs with similar temperature and metal abundance as \target\ see fig.\,1 of \citet{xuetal19-1} and fig.\,1 of \citet{wilsonetal15-1}.}, has a measurable effect. Ultraviolet line blanketing in ``warm'' DB[A]Z white dwarfs has not been systematically explored (for a discussion see \citealt{dufouretal12-1} and \citealt{coutuetal19-1}), and is beyond the scope of this paper. We proceed here with an iterative approach to optimise all atmospheric parameters: \Teff, \logg, \htohe, and the individual metal abundances. In what follows, we briefly outline the overall strategy and describe the details of the atmosphere models and the methods that we used.


Using a grid of He+H model spectra, we first determined \Teff, \logg\ and \htohe\ from the WHT average spectrum (Section~\ref{sec:justspec}). We also fit the available broad-band photometric points to obtain an independent measurement of \Teff\ and \logg, making use of the \textit{Gaia} Data Release 2 (DR2) parallax as an additional constraint, and fixing \htohe\ to the spectroscopic value. Note that this is necessary, as the photometry alone is not sensitive to the chemical abundance (Section~\ref{sec:justphot}). We found that the results from both fits disagreed and therefore implemented a hybrid approach that provided the parameters that best fit the spectroscopy and the photometry simultaneously (Section~\ref{sec:hybridapp}).

In order to obtain the metal abundances we produced a grid of synthetic spectra for each metal detected, with \Teff, \logg, and \htohe\ fixed at the values estimated above. The abundances were determined by fitting the metal absorption lines in the WHT and Keck spectra (Section~\ref{sec:metal_abs}).

With these initial metal abundances fixed, we produced a new grid of synthetic He+H+Z spectra and refine the values of \Teff, \logg, and \htohe\ with the spectroscopic, photometric and hybrid approaches. We then iterate over this whole process using new sets of He+H+Z model spectra in each step until convergence is achieved.

\subsection{\label{sec:models}Atmosphere models and fitting procedure}
The synthetic spectra were computed using the code of \citet{koester10-1}. DB white dwarfs have substantial outer CVZs, which are treated in the model atmospheres following a 1D approach that includes the mixing length, $\mathrm{ML2}/\alpha$. Past studies have used $\mathrm{ML2}/\alpha$ in the range 0.6 to 1.25, with higher values used for helium-dominated atmospheres \citep{vossetal07-1, bergeronetal11-1}. In a study of pure helium atmospheres, \citet{cukanovaiteetal18-1} showed that the spectroscopic analysis of DBs using a full 3D treatment of the CVZ results in somewhat different atmospheric parameters compared to the 1D models. The situation complicates further if the envelope contains traces of hydrogen, and \citet{cukanovaiteetal19-1} derived a parameter-dependent empirical calibration of the mixing length parameter from a suite of 3D models sampling a wide range in \Teff, \logg, and \htohe. They found $\mathrm{ML2}/\alpha = 1.0$ to be the best approach within the parameters of \target\ \citep[see fig. 10 and 11 in][]{cukanovaiteetal19-1}. Our initial He+H grid of model spectra spanned $T_\mathrm{eff}=10\,000-30\,000$\,K in steps of $200$\,K,  $\log g=7.0-9.0$ in steps of 0.2\,dex, and $\htohe=-5.0$ to 0.0 in steps of 0.2\,dex and had no metals included. For the subsequent iterations, we produced grids of He+H+Z model spectra exploring the same \Teff, \logg, and \htohe\ parameter space, but now including metals with the abundances estimated in the previous iteration.

We used the Markov-Chain Monte Carlo (MCMC) {\sc emcee} package within {\sc python} \citep{foreman13} to fit the model spectra to the observed data. We explored the parameter space and minimised the corresponding $\chi^{2}$ using 150 different seeds and $40\,000$ steps per seed. We employed flat priors for all the parameters except for the parallax, $\varpi$, for which we used a Gaussian prior. The MCMC analysis provides statistical uncertainties on all parameters and reveals correlations between them. In addition, we carried out a more realistic analysis of the uncertainties also taking into account systematic errors (hereafter quoted in brackets next to the statistical uncertainties).

\subsection{Spectral fitting} \label{sec:justspec}
We derived \Teff, \logg, and \htohe\ by fitting our grid of model spectra to the WHT average spectrum. The synthetic spectra were first degraded to the FWHM resolution of the observed data (1.9\,\AA). We then normalized the helium and Balmer absorption lines in both the observed and model spectra by using low-order polynomial fits to the surrounding continuum regions. Hence, the spectral fitting relied entirely on the shape of the Balmer and helium absorption line profiles \citep[e.g.][]{koesteretal81-1, bergeronetal92-1}. Some of these profiles have metal lines superimposed that were masked out for the fitting process. However, the Mg and Si lines between \Line{He}{i}{3820} and \Line{He}{i}{3889}, and those bracketing \Line{He}{i}{5048}, were included in the analysis since their removal would have left too little of the helium line profiles available for a reliable fit (see Fig.~\ref{fig:htohefit}). 

We note that we made no attempt to determine the photospheric parameters from the Keck spectrum, as most of the Balmer and helium lines extend over more than one spectral order, which could not be reliably flux calibrated due to the lack of a spectrophotometric standard star obtained alongside the observations of \target. 

The best-fit model was found for $T_\mathrm{eff}=16\,417 \pm 55\,(95)$\,K, $\log g=8.14 \pm 0.01\,(0.02)$\,dex, and $\htohe=-3.68 \pm 0.02\,(0.04)$\,dex (Table~\ref{tab:AP_comp}). Notice that this model includes the metal abundances provided in Table~\ref{tab:metalabs}. We estimated the systematic uncertainties from fitting ten different subsets of helium and Balmer lines in the same MCMC fashion mentioned above, and computing the standard deviation, where we ensured that each subset included lines of both elements. 

While the spectral modelling alone provided good constraints on the three atmospheric parameters (see Fig.\,\ref{fig:corner_plots}, left panel), neutral helium transitions are less sensitive to effective temperature and surface gravity changes than their hydrogen counterparts.

\subsection{Photometric fitting}\label{sec:justphot}

To estimate the atmospheric parameters from photometric data the synthetic spectra need to be scaled with the solid angle of the star, $\pi \times (\Rwd/D)^{2}$, with $D=1000/\varpi$ ($\varpi$ is the parallax in mas) the distance to the source and $\Rwd$ its radius. Besides, photometric data are affected by interstellar extinction, so we reddened the models by $E(B-V)=0.01$, as determined from the 3D dust map produced by Stilism\footnote{\url{https://stilism.obspm.fr/}} and the distance to \target\ ($D=91.15$\,pc) derived from its \textit{Gaia} Data Release 2 \citep[DR2,][]{gaiaDR2-1} parallax ($\varpi=10.97 \pm 0.05$\,mas). Scaling and reddening place the grid of synthetic spectra on an absolute flux scale that can then be compared with existing photometric data (see Table~\ref{tab:phot_points}). The photometric magnitudes were converted into fluxes using the appropriate zero-points available on the Spanish Virtual Observatory (SVO) Filter Profile Service\footnote{\url{http://svo2.cab.inta-csic.es/theory/fps/}}, while synthetic model fluxes were computed for each photometric bandpass. $\Rwd$ is a function of \Teff\ and \logg\ and was obtained using the mass-radius relation of \citet{fontaineetal01-1} for C/O cores and thin hydrogen layers ($10^{-10} M_{\mathrm{H}}/\Mwd$, with $M_{\mathrm{H}}$ and \Mwd\ the hydrogen and white dwarf mass, respectively). The photometric fit is not sensitive either to \htohe\ or the metal abundances, but as mentioned above, they affect the structure of the white dwarf photosphere. Hence, these abundances were fixed as determined from the spectroscopic and metal fit in each iteration. We then fitted the three free parameters \Teff, \logg, and $\varpi$ again using the MCMC procedure outlined above, using flat priors for \Teff\ and \logg, and a Gaussian prior for $\varpi$.

\begin{figure}
\centering
	\includegraphics[width=1.04\columnwidth]{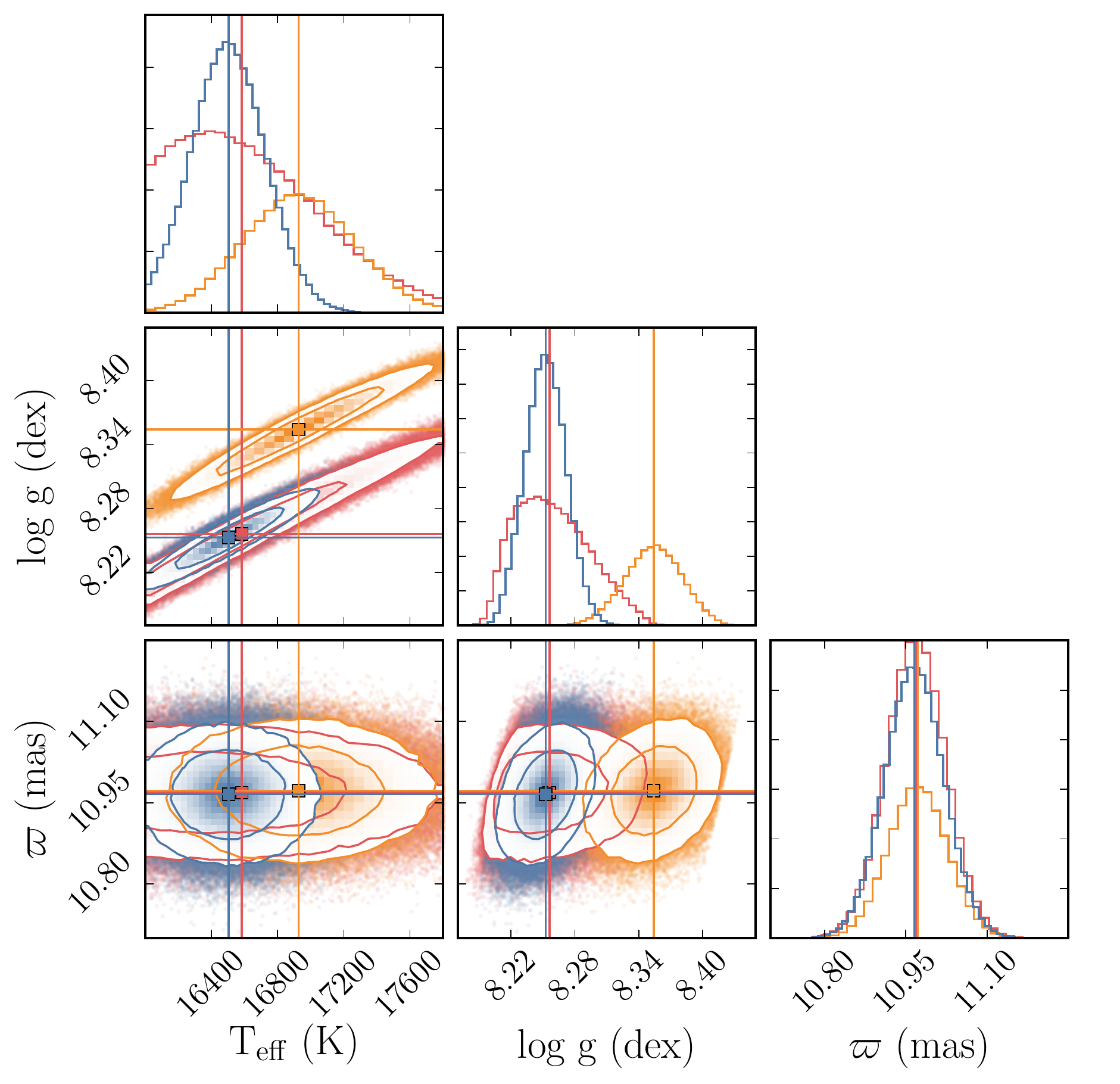}
    \caption{Probability distributions of \Teff, \logg, and $\varpi$\ derived from three different photometric catalogues: SDSS (red), Pan--STARRS1 (blue) and \textit{Gaia} DR2 (orange). Note that we used the three \textit{Gaia} filters for this plot: $G_\mathrm{BP}$, $G$ and $G_\mathrm{RP}$. The vertical solid lines mark the median values and the contour lines identify the 1- and 2-$\sigma$ regions.}
    \label{fig:phot_corner}
\end{figure}

We fitted the SDSS, the Panoramic Survey Telescope and Rapid Response System (Pan--STARRS1) Data Release~1 (DR1) and the \textit{Gaia} DR2 photometry\footnote{Additional broad-band photometry of \target\ is available, but was not included in the fit for the following reasons: The Galaxy Evolution Explorer (\textit{GALEX}) ultraviolet fluxes are likely to be affected by blanketing from metal lines, the AAVSO Photometric All-Sky Survey (APASS) photometry quality is inferior to the other optical data already used, and the Two Micron All-Sky Survey (2MASS) and Wide-field Infrared Survey Explorer (\textit{WISE}) infrared observations do not add any useful constraints to the atmospheric parameters.} separately. The best-fit parameters and their statistical uncertainties are listed in Table\,\ref{tab:AP_comp}, and the probability distributions of the parameters are shown in Fig.\,\ref{fig:phot_corner}. We then combined the three photometric data sets and repeated the fitting process, now excluding the extremely broad-band \textit{Gaia} $G$-band, which encompasses all other bands and hence does not add in terms of useful constraints. We obtained $T_\mathrm{eff}= 16\,913 \pm 170\,(285) $ K, $\log g= 8.28 \pm 0.01\,(0.02)$\,dex, and $\varpi = 10.97 \pm 0.05\,(0.06)$ mas as best-fit parameters for the combined photometry. We note that this fit was carried out with synthetic spectra that included the metal abundances provided in Table~\ref{tab:metalabs}. The systematic uncertainties were estimated as the average of the residuals obtained by computing ten different MCMC fits with different combinations of the photometric bands drawn from the three individual sets.

The \Teff\ provided by the photometric fit is 500\,K hotter than that obtained using the spectroscopic method (Table\,\ref{tab:AP_comp}). The difference in \logg\ is significant ($\simeq 0.14$\,dex), but the width and depth of the helium transitions are not very sensitive to changes in \logg. The derived parallax is entirely compatible with the \textit{Gaia} DR2 value.

\begin{table}
\caption{Photometry of \target. We list the point spread function (PSF) SDSS magnitudes \citep{SDSSpass} and the mean PSF Pan--STARRS1 magnitudes \citep{PanSTARRS2}.}
%
\begin{tabular}{ccc}
\hline
\textit{Gaia} DR2  & Pan--STARRS1 DR1 & SDSS \\
(Vega mag) & (AB mag) & (AB mag) \\
\hline\noalign{\smallskip}

$G_\mathrm{BP} = 16.093 \pm 0.005$          &                            & $u=15.893 \pm 0.032$   \\
$G_\mathrm{~~\,\,} = 16.223 \pm 0.001$      & $g = 16.133 \pm 0.006 $    & $g=16.106 \pm 0.018$   \\
$G_\mathrm{RP} = 16.313 \pm 0.005$          & $r = 16.374 \pm 0.002 $    & $r=16.377 \pm 0.012$   \\
                                            & $i = 16.642 \pm 0.003$     & $i=16.612 \pm 0.012$   \\
                                            & $z = 16.886 \pm 0.006$     & $z=16.912 \pm 0.019$   \\
                                            & $y = 17.042 \pm 0.007$     &                        \\
\noalign{\smallskip}\hline
\end{tabular}
\label{tab:phot_points}
\end{table}

\subsection{Hybrid fitting}
\label{sec:hybridapp}

The discrepant results obtained with the two different methods above motivated us to explore a hybrid technique which combines the spectroscopic sensitivity to \htohe\ with that to \Teff\ and \logg\ of the photometric method. As before, we used an MCMC optimisation to fit the model spectra to the WHT spectroscopy (we used the same absorption lines as in Section~\ref{sec:justspec}, see Fig.~\ref{fig:htohefit}) and to the SDSS, Pan--STARRS1 and \textit{Gaia} photometry (again excluding the broad \textit{Gaia} $G$-band). We minimized the sum of the spectroscopic and photometric $\chi^2$ values using four free parameters: \Teff, \logg, \htohe, and $\varpi$. We adopted flat priors except for $\varpi$, for which we used a Gaussian prior. 
 
\begin{figure*}
\begin{minipage}[t][4.5cm][t]{.42\textwidth}
\centering
	\includegraphics[width=1\columnwidth]{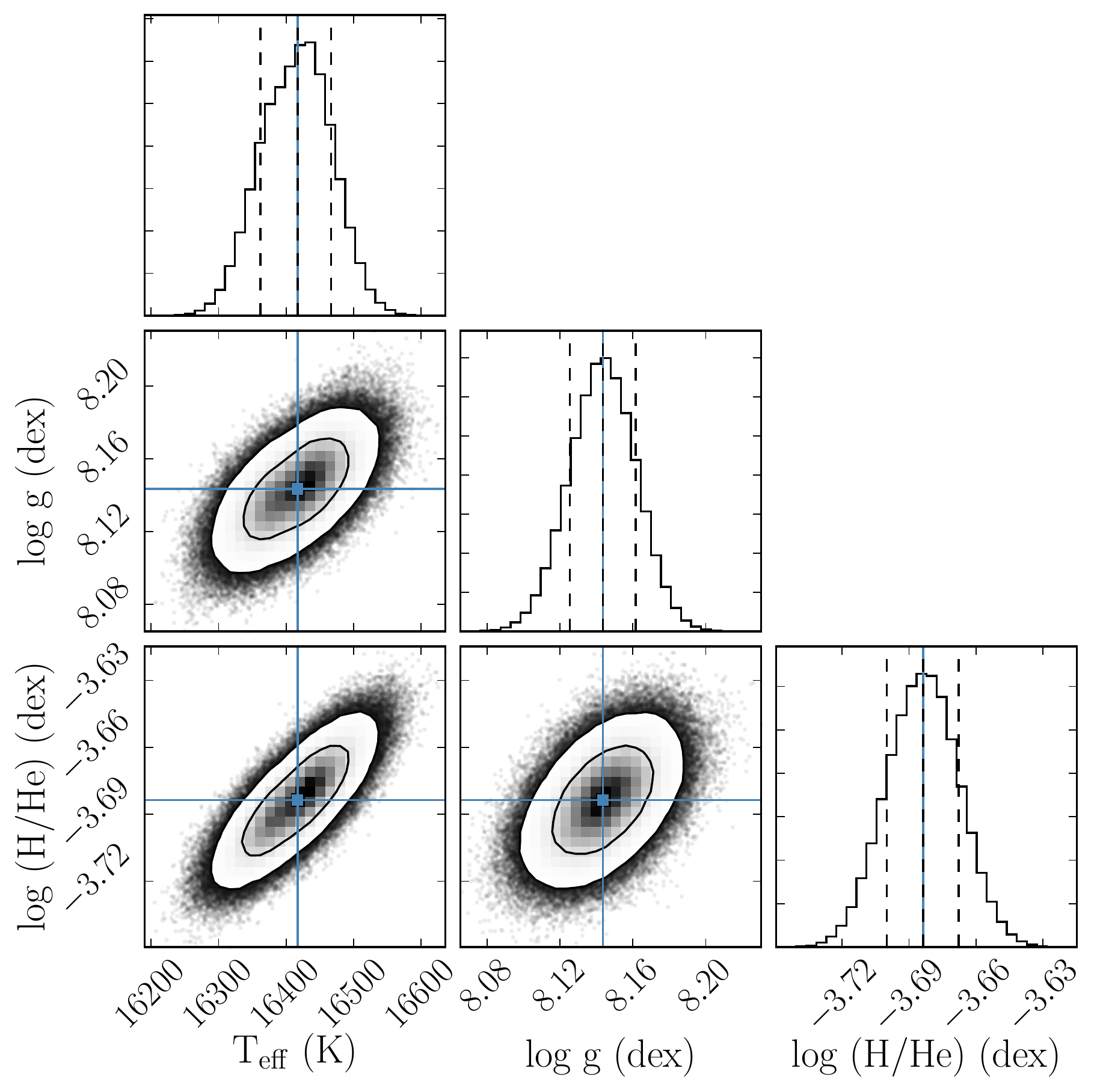}
\end{minipage}
\begin{minipage}[t][2.18cm][b]{.532\textwidth}
\includegraphics[width=1.025\columnwidth]{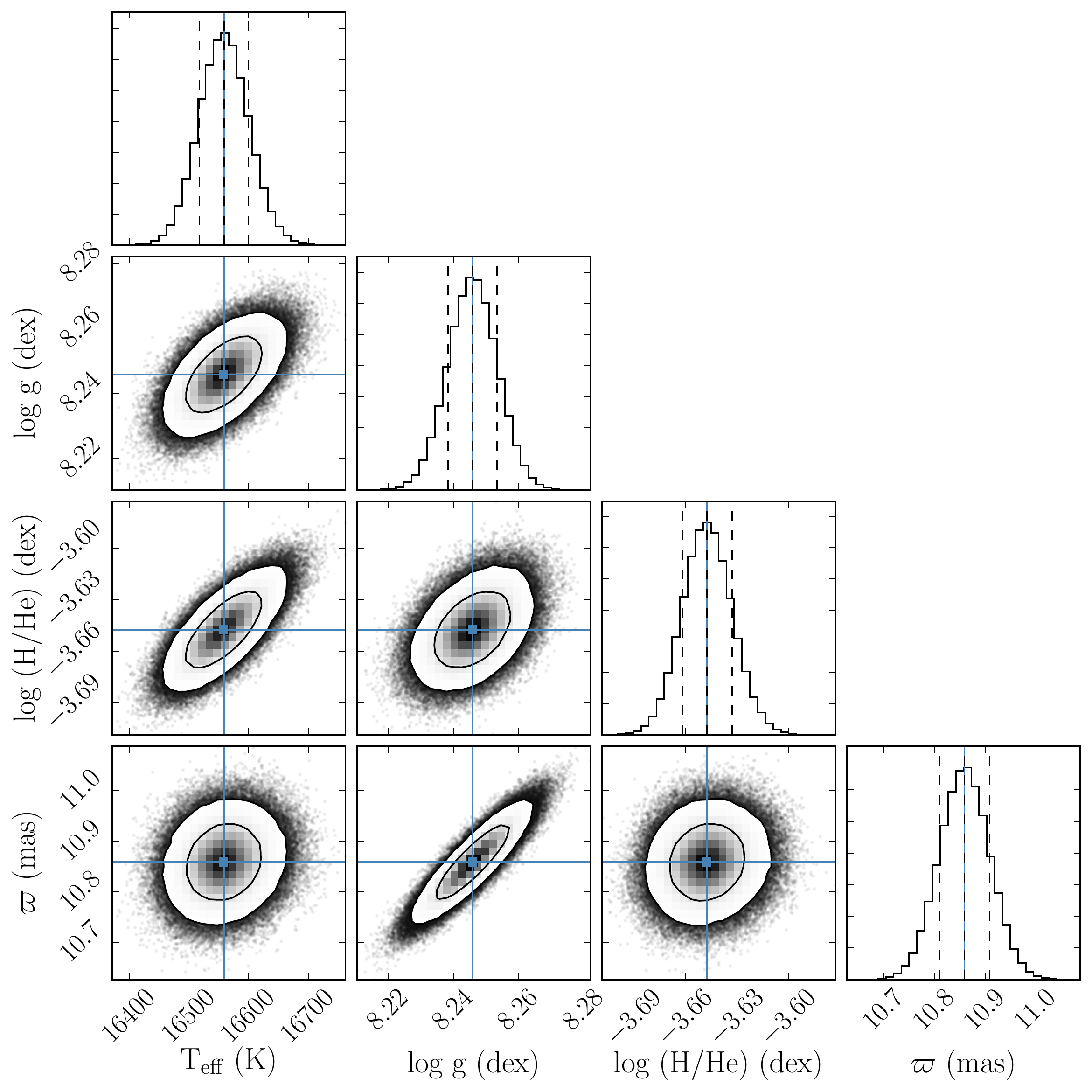}
\end{minipage}
\caption{Probability distributions of the photospheric parameters derived from the spectroscopic (left) and hybrid (right) fits. The vertical solid line marks the 0.5 quantile, and the vertical dashed lines the 0.16 and 0.84 quantiles. The contour lines identify the $1$- and 2-$\sigma$ regions. We note that just the statistical uncertainties derived from the MCMC are shown in this plot.}
\label{fig:corner_plots}
\end{figure*}

\begin{figure*}
\centering
	\includegraphics[width=2.11\columnwidth]{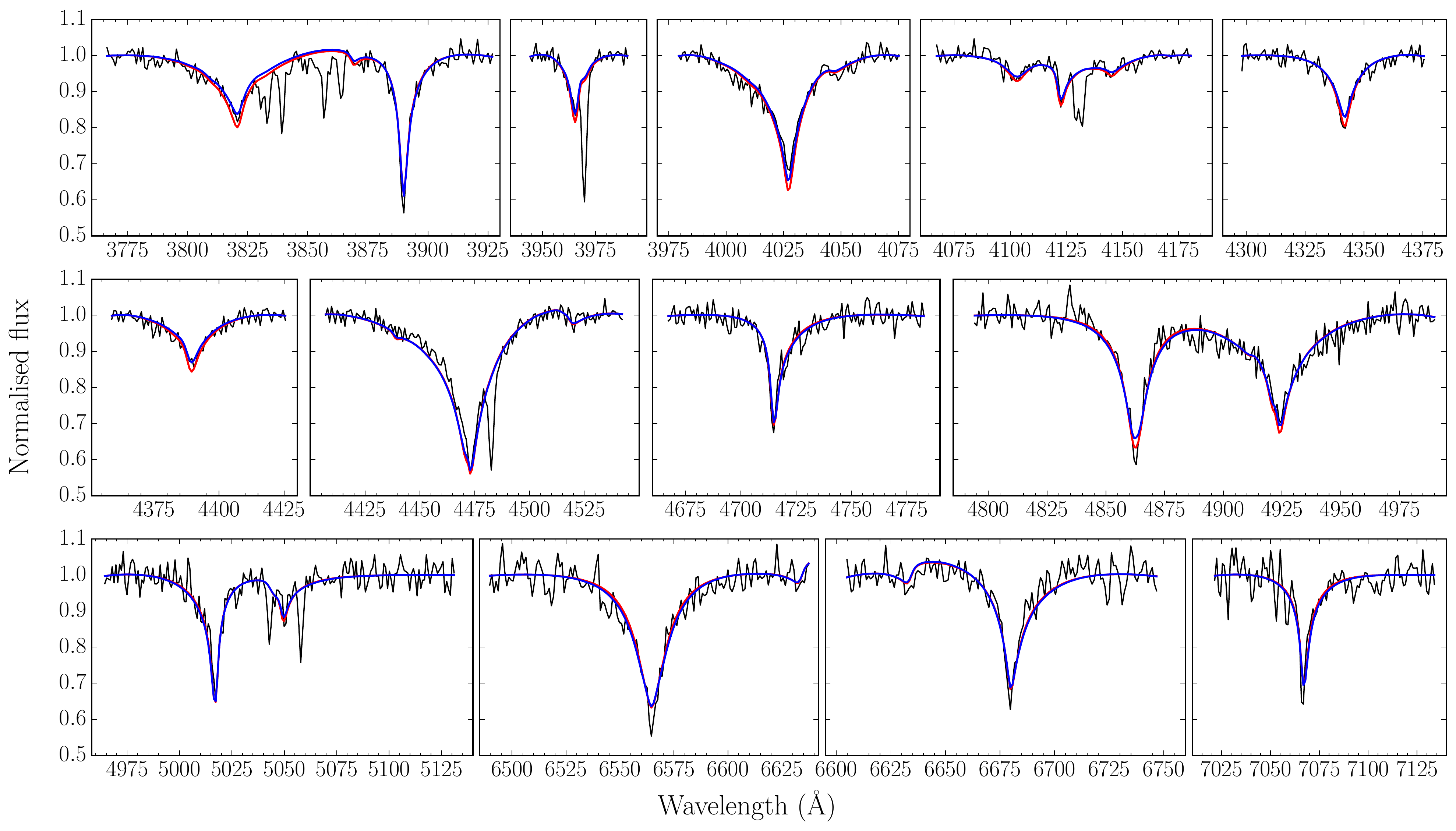}
    \caption{Hydrogen and helium absorption lines used in the spectroscopic and hybrid fits to the WHT average spectrum. The best-fit models for each approach are over-plotted in red and blue, respectively. See Table~\ref{tab:AP_comp} for details.}
    \label{fig:htohefit}
\end{figure*}

The best-fit parameters are: $T_\mathrm{eff}=16\,560 \pm 42\,(75)$\,K, $\log g=8.25 \pm 0.01\,(0.02)$\,dex, $\htohe=-3.65 \pm 0.01\,(0.03)$\,dex, and $\varpi = 10.86 \pm 0.05\,(0.05)$\,mas (see right panel of Fig.~\ref{fig:corner_plots}). The systematic errors were estimated by performing several MCMC fits with different subsets of photometric points and helium and Balmer absorption lines. Based on white dwarf cooling models\footnote{\url{http://www.astro.umontreal.ca/~bergeron/CoolingModels} \citep{fontaineetal01-1,holberg+bergeron06-1, kowalski+saumon06-1, tremblayetal11-2}}, these correspond to a white dwarf mass, radius, and cooling age of $\Mwd=0.77\pm0.01\,\Msun$, $\Rwd=0.0109\pm0.0001\,\Rsun$, and $\tau_\mathrm{cool}=215\pm10$\,Myr, respectively.

The absorption lines used in both the spectroscopic and hybrid approaches are displayed in Fig.~\ref{fig:htohefit}, with the best-fit models over-plotted in blue and red, respectively. The spectroscopic and hybrid solutions led to a difference in \Teff\ and \logg\ of $140$\,K and $\simeq 0.1$\,dex, that we interpret as the result of the higher \logg\ forced by the inclusion of the parallax as a free parameter. However, the line profiles of the two models are nearly indistinguishable, illustrating that both sets of solutions are consistent with the WHT spectrum. The $\varpi$ derived from the hybrid method is $0.11$\,mas smaller than the \textit{Gaia} DR2 value\footnote{We note that the \textit{Gaia}~eDR3 value of the parallax is $10.89\pm0.04$\,mas, which is in good agreement with the result from the hybrid fit. However, we refrain from repeating the analysis in the light of eDR3 \citep{lindegren20}, as it would not affect any of the further discussions and conclusions in this paper.}. This discrepancy is very likely related to the lower \logg\ of the hybrid fit as compared to the photometric value (see Table\,\ref{tab:AP_comp}), that arises from the inclusion of the spectroscopy, which forces the hybrid fit to settle on a slightly larger distance to compensate.

The discrepancies between the photometric, spectroscopic and hybrid fits underline that there remain systematic uncertainties within the atmospheric models, the observational data, and the different fitting methods. Further exploration of these uncertainties is necessary, but cannot be within the context of this study of an individual star. We will carry out a detailed comparison of the different fitting procedures using a larger sample of DBA white dwarfs in a forthcoming paper. For the remaining analysis and discussion, we will adopt the atmospheric parameters of \target\ resulting from the hybrid analysis, that are derived from fitting all the free parameters simultaneously to the available data. We note that the small differences in the atmospheric parameters between the different fitting approaches have no impact on either the abundance analysis carried out in Section~\ref{sec:metal_abs} or its interpretation.

\begin{table}
\centering
\caption{Atmospheric parameters of \target\ obtained through the spectroscopic, photometric and hybrid fitting techniques. The photometric fits account for the final photometric analysis (``Phot'', all filters included except \textit{Gaia} $G$; see text for details) and the three individual photometric data sets: SDSS, Pan--STARRS1 (PS1) and \textit{Gaia}. The \htohe\ value was fixed to the spectroscopic best-fit value. We note that all the uncertainties quoted here also account for systematic errors, except for the data sets labelled with $^\star$, that indicate pure statistical errors derived from the MCMC fitting.}
\begin{tabular}{lcccc}
\hline
Fitting & $T_{\mathrm{eff}}$  & log\,$g$  & $\log\,(\mathrm{H/He})$  & $\varpi$ \\
 & (K) & (dex) & (dex) & (mas) \\
\hline\noalign{\smallskip}
Spec & $16\,417 \pm 95$ & 8.14 $\pm$ 0.02 & $-3.68 \pm 0.04$ & - \\
\hline
\vspace{0.15cm}
Phot & $16\,913 \pm 285$ & $8.28 \pm 0.02$ & --3.68 & $10.97 \pm 0.06$ \\
\vspace{0.15cm}
SDSS$^{\star}$ & $16\,580^{+505}_{-380}$ & $8.26^{+0.04}_{-0.03}$ & --3.68 & $10.97 \pm 0.05$ \\
\vspace{0.15cm}
PS1$^{\star}$ & $16\,505 \pm 225$ & $8.25 \pm 0.02$ & --3.68 & $10.96 \pm 0.05$ \\
\vspace{0.15cm}
\textit{Gaia}$^{\star}$ & $16\,926 \pm 335$ & $8.35 \pm 0.03$ & --3.68 & $10.97 ^{+0.06}_{-0.05}$ \\
\hline
Hybrid & $16\,560 \pm 75$ & $8.25 \pm 0.02$ & $-3.65 \pm 0.03$ & $10.86 \pm 0.05$ \\
\noalign{\smallskip}\hline
\end{tabular}
\label{tab:AP_comp}
\end{table}

\subsection{\label{sec:metal_abs} Photospheric metal abundances}
The WHT spectrum of \target\ presents metal absorption lines of oxygen, magnesium, silicon and calcium. In addition, we also identified traces of sodium, aluminium, titanium, chromium, manganese, iron and nickel in the Keck spectrum. This is mainly a consequence of the much superior spectral resolution of the Keck data ($0.2$\,\AA) compared with the WHT. Therefore, we used the Keck spectrum to derive all the metal abundances except for oxygen, whose absorption lines were only covered by the WHT spectrum. However, and for comparison purposes, we also measured the abundances of magnesium, silicon and calcium in the WHT spectrum, using lines not necessarily common to the Keck spectrum. Keeping \Teff, \logg\ and \htohe\ fixed to the values obtained from the hybrid fitting technique in each step of the iteration (Section~\ref{sec:hybridapp}), we generated grids of model spectra for each of the metals (Z) identified. We explored the $\log(\mathrm{Z}/\mathrm{He})$ space between $-9.0$ and $-3.0$ in steps of $0.1$ dex, and followed the same procedure as in Section~\ref{sec:justspec} to continuum-normalize the regions around each of the metal absorption lines to be included in the fit (see Table~\ref{tab:lines}), degraded the models to the corresponding instrument spectral resolution, and carried out the $\chi^2$ minimization via MCMC independently for each metal.

\begin{figure*}
\centering
\includegraphics[width=2.1\columnwidth]{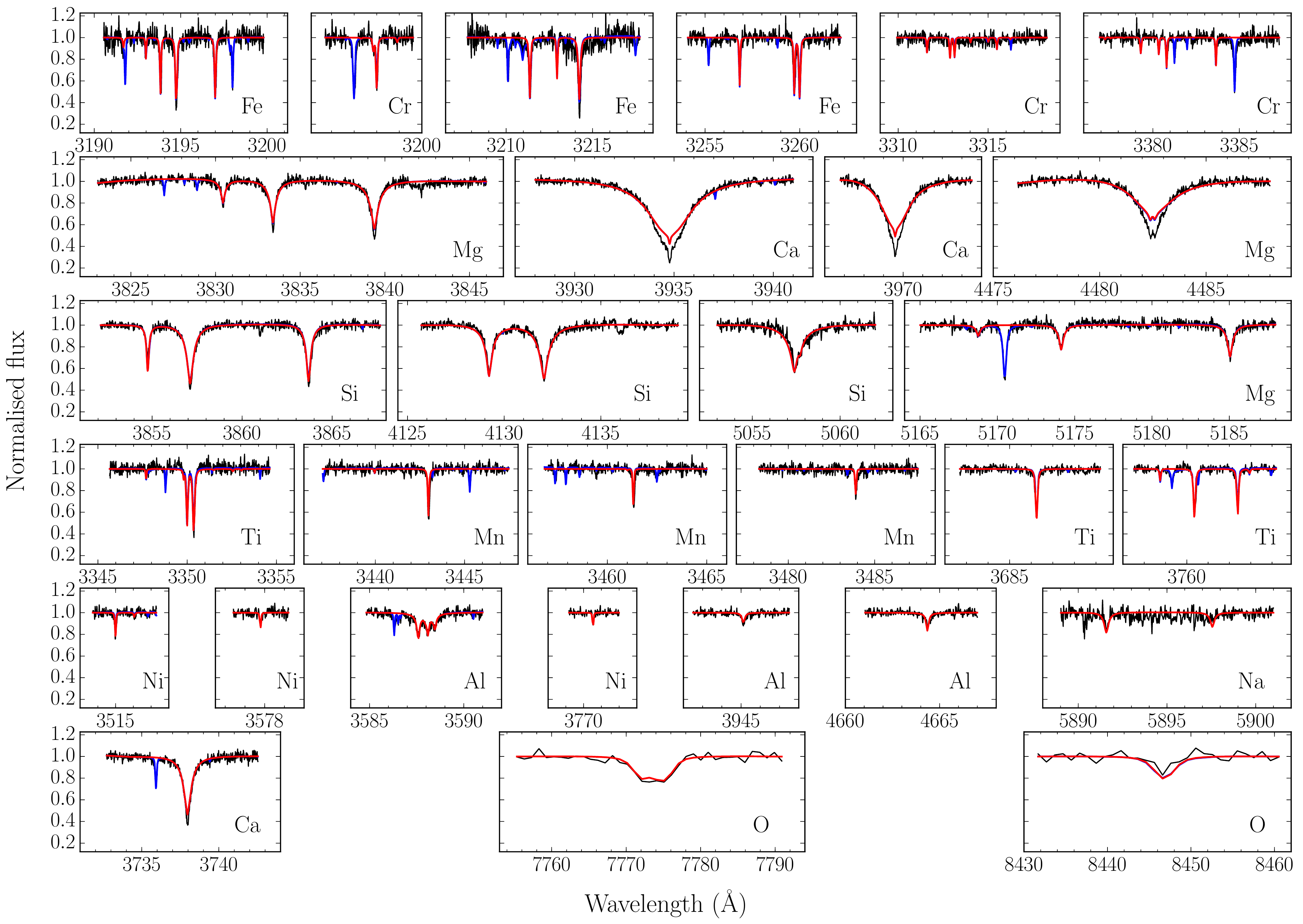}
\caption{Some of the major absorption lines used to derive the photospheric metal abundances from the Keck spectrum. The exception are the oxygen lines that were only recorded in the WHT spectrum. The best-fit model including all the metals found in the spectra and only the element labelled in each panel are over-plotted in blue and red, respectively.}
\label{fig:metal_lines}
\end{figure*}

The metal abundances derived for the WHT and Keck spectra are listed in Table~\ref{tab:metalabs}. We estimated the systematic errors by repeating the spectral fitting for each metal with different subsets of lines within the same spectrum. Fig.~\ref{fig:metal_lines} displays the best-fit models over-plotted on some of the lines used to measure the abundances. We note that the only transition of \Ion{Na}{i} is the 5890/96\,\AA\ absorption doublet, that is only marginally detected and hence we only quoted an upper limit on its abundance.

\begin{table}
\caption{Spectral lines used in the determination of the metal chemical abundances.}
\begin{tabular}{ll}
\hline
Ion & Air wavelength (\AA)\\
\hline
O\,{\sc i} & 7771.94, 7774.17, 7775.39, 8446.36\\
Na\,{\sc i} & 5889.95 , 5895.92\\
Mg\,{\sc i} & 3332.15, 3336.67, 3829.36, 3832.30, 3838.29, 5167.32, 5172.68,\\
            & 5183.60\\
Mg\,{\sc ii} & 3838.29, 3850.39, 4384.64, 4390.56, 4427.99, 4433.99, 4481.13,\\
            & 4481.33\\
Al\,{\sc i} & 3944.01\\
Al\,{\sc ii} & 3586.56, 3587.07, 3587.45, 4663.06\\
Si\,{\sc ii} & 3853.66, 3856.02, 3862.60, 4128.07, 4130.89, 5055.98\\
Ca\,{\sc i} & 4226.73 \\
Ca\,{\sc ii} & 3158.87, 3179.33, 3181.28, 3706.03, 3736.90, 3933.66, 3968.47 \\
Ti\,{\sc ii} & 3168.52, 3190.88, 3202.53, 3217.05, 3218.27, 3222.84, 3224.24,\\
           & 3228.61, 3229.19, 3229.42, 3232.28, 3236.57, 3239.04, 3239.66,\\
           & 3241.98, 3248.60, 3251.91, 3252.91, 3254.25, 3261.58, 3278.29,\\
           & 3278.92, 3287.65, 3321.70, 3322.94, 3329.45, 3332.11, 3335.19,\\
           & 3341.87, 3349.03, 3349.40, 3361.21, 3372.79, 3380.28, 3383.76,\\
           & 3387.83, 3394.57, 3444.31, 3504.89, 3510.84, 3535.41, 3685.19,\\
           & 3741.64, 3759.29, 3761.32, 4163.64, 4171.90, 4290.22, 4294.09,\\
            & 4300.04, 4549.62,  4571.97\\
Cr\,{\sc ii} & 3180.70, 3183.33, 3196.92, 3197.08, 3209.18, 3216.55, 3217.40,\\
            & 3234.06, 3238.76, 3269.08, 3269.76, 3270.13, 3291.76, 3295.43,\\ 
            & 3310.66, 3311.93, 3312.18, 3314.54, 3324.06, 3324.13, 3324.34,\\  
            & 3342.58, 3378.33, 3379.37, 3379.82, 3382.68, 3402.40, 3403.32,\\
            & 3408.77, 3421.21, 3422.74, 3585.29, 3585.50, 3677.68, 3677.84\\
Mn\,{\sc ii} & 3441.98, 3460.31, 3474.04, 3474.13, 3482.90, 3495.83, 3496.81,\\
           &  3497.53\\
Fe\,{\sc i} & 3190.82, 3211.99, 3227.79, 3249.50\\  
Fe\,{\sc ii} & 3135.36, 3144.75, 3167.86, 3170.34, 3177.53, 3183.07, 3185.32,\\
            & 3186.74, 3187.30, 3192.07, 3192.91, 3193.80, 3196.07, 3210.45,\\
            & 3213.31, 3227.74, 3231.71, 3232.79, 3236.83, 3237.40, 3237.82,\\
            & 3243.72, 3247.18, 3247.39, 3255.87, 3258.77, 3259.05, 3266.94,\\ 
            & 3276.60 ,3277.35, 3281.29, 3289.35, 3295.82, 3297.88, 3323.06,\\
            & 3436.11, 3456.93, 3468.68, 3493.47, 3621.27, 3624.89, 3748.49,\\
            & 4233.16, 4303.17, 4351.76, 4508.28, 4522.63, 4549.20, 4549.47,\\
            & 4555.89, 4583.83, 5197.57, 5216.86, 5227.49, 5234.62, 5275.99,\\ 
            & 5316.61\\
Ni\,{\sc i} & 3414.76, 3465.6, 3471.3, 3513.93, 3515.05, 3524.54, 3576.73, \\
            & 3769.24\\
\noalign{\smallskip}\hline
\end{tabular}
\label{tab:lines}
\end{table}

\begin{table}
\caption{Element photospheric abundances derived from the Keck and the WHT spectra (hybrid best-fit parameters: $\Teff=16\,560$~K, $\logg=8.25$~dex and $\htohe =-3.65$~dex), diffusion velocities, and sinking times for each element in the CVZ.}
\centering
\begin{tabular}{clcc}
\hline
Element &\hspace{1.1cm}log\,($Z$/He)  & $v_\mathrm{z}$  & $\tau_\mathrm{z}$  \\
&\hspace{1.3cm}(dex) & ($10^{-7}$ cm s$^{-1}$)& ($10^{4}$ yr)\\
&\hspace{0.69cm}Keck\hspace{0.31cm}/\hspace{0.2cm}WHT & & \\
\hline
\noalign{\smallskip}
H  & \hspace{0.6cm}-\hspace{0.85cm} /\hspace{0.12cm}$-3.65 \pm 0.03$ & - & - \\ 
\vspace{0.1cm}
O  & \hspace{0.6cm}-\hspace{0.85cm} /\hspace{0.12cm}$-4.59 \pm 0.12$ & 2.30 & 12.20\\
\vspace{0.1cm}
Na &\hspace{0.26cm} $<-$6.5\hspace{0.55cm}/\hspace{0.9cm}-                   & 3.10 & 9.06\\
\vspace{0.1cm}
Mg & $-5.15 \pm 0.04$\hspace{0.12cm}/\hspace{0.14cm}$-5.18 \pm 0.02$          & 3.09 & 9.07\\
\vspace{0.1cm}
Al & $-6.3 \pm0.1$\hspace{0.12cm}/\hspace{0.9cm}-               & 3.41 & 8.22\\
\vspace{0.1cm}
Si & $-5.29 \pm 0.04$\hspace{0.12cm}/\hspace{0.14cm}$-5.36\pm 0.02$          & 3.38 & 8.30\\
\vspace{0.1cm}
Ca & $-6.15 \pm 0.05$\hspace{0.12cm}/\hspace{0.14cm}$-6.05\pm 0.02$          & 4.34 & 6.47\\
\vspace{0.1cm}
Ti & $-7.78^{+0.08}_{-0.09}$\hspace{0.35cm}/\hspace{0.9cm}-         & 5.26 & 5.34\\
\vspace{0.1cm}
Cr & $-7.19^{+0.07}_{-0.06}$\hspace{0.35cm}/\hspace{0.9cm}-         & 5.54 & 5.06\\
\vspace{0.1cm}
Mn & $-7.64^{+0.06}_{-0.07}$\hspace{0.35cm}/\hspace{0.9cm}-         & 5.82 & 4.82\\
\vspace{0.1cm}
Fe & $-5.53^{+0.10}_{-0.12}$\hspace{0.35cm}/\hspace{0.9cm}-         & 5.79 & 4.84\\
Ni &$-6.93\pm0.10$\hspace{0.12cm}/\hspace{0.9cm}-                & 5.90 & 4.75\\
\noalign{\smallskip}\hline
\end{tabular}
\label{tab:metalabs}
\end{table}

\section{Discussion}
\label{sec:discussion}
\subsection{Chemical abundances of the parent body}
\label{sec:PB_abs}

In the previous section we estimated the photospheric abundances relative to helium of the 11 metals identified in our spectra. Translating these abundances into the bulk composition of the parent body needs taking into account the diffusion velocities of the individual chemical elements and the accretion history of \target. 

Following \citet{koester09-1}, we assume a simple scenario for the accretion history in which the mass transfer to the white dwarf switches on, remains at a constant rate for some time, and then switches off\,\footnote{Reality is likely more complicated, with a time-dependent variation of the mass accretion rate.}. In this scenario, accreting white dwarfs can be found in three different phases: the increasing state, the steady state, and the decreasing state. In the increasing state, planetary debris is accreted on to an initially pure-helium atmosphere, and metal abundances increase linearly with time. Early in the increasing state, the photospheric abundances are approximately those of the parent body, but they begin to diverge later on because the individual elements settle out of the CVZ with different velocities. In the steady state, an accretion/diffusion equilibrium is reached, all elements diffuse out of the CVZ (with their own individual velocities) at the same rate as they are accreted, and the parent body composition can be reliably derived. In the decreasing phase, the accretion has stopped and the abundances of the accreted elements decline exponentially and proportional to their individual sinking time-scales.

Determining the accretion state of metal-polluted DBs is highly uncertain as the diffusion time-scales are of the order of the estimated lifetimes of the circumstellar discs \citep[$10^{4}-10^{6}$\,yr, ][]{girven12}. Consequently, reaching the steady state can take $\sim10^{5}-10^{6}$\,yr. Similarly, the accretion/diffusion equilibrium can only last for a few diffusion times due to the limited accretion disc lifetime. Finally, once accretion stops and the system enters the decreasing state, photospheric metals will remain detectable for $\sim10^5-10^6$\,yr. 

Assuming the accretion rate, $\dot M_{\mathrm{z}}$, diffusion velocity, $v_{\mathrm{z}}$, and settling time, $\tau_{\mathrm{z}}$ as constants \citep{koester09-1}, the time-dependent mass abundance, $Z_\mathrm{ph,m}$, of each metal in the CVZ (which is identical to the photospheric mass abundance because of the rapid mixing in the CVZ) is given by:
\begin{equation}
Z_{\mathrm{ph,m}} (t) = Z_{\mathrm{ph,m}}(0)\,e^{-t/\tau_{\mathrm{z}}} + \frac{\tau_{\mathrm{z}}\dot{M_{\mathrm{z}}}}{M_{\mathrm{CVZ}}}\left[1-e^{-t/\tau_{\mathrm{z}}}\right]~,
\label{eq:mass_ab}
\end{equation}

\noindent
with $M_\mathrm{CVZ}$ the mass contained in the CVZ. The parent body composition is mirrored in the accretion rates of the individual elements, $\dot{M_{\mathrm{z}}}$, and can be computed from the measured photospheric abundances given an assumption of the time $t$ at which the system is observed.

\begin{figure}
\centering
\includegraphics[width=1.0\columnwidth]{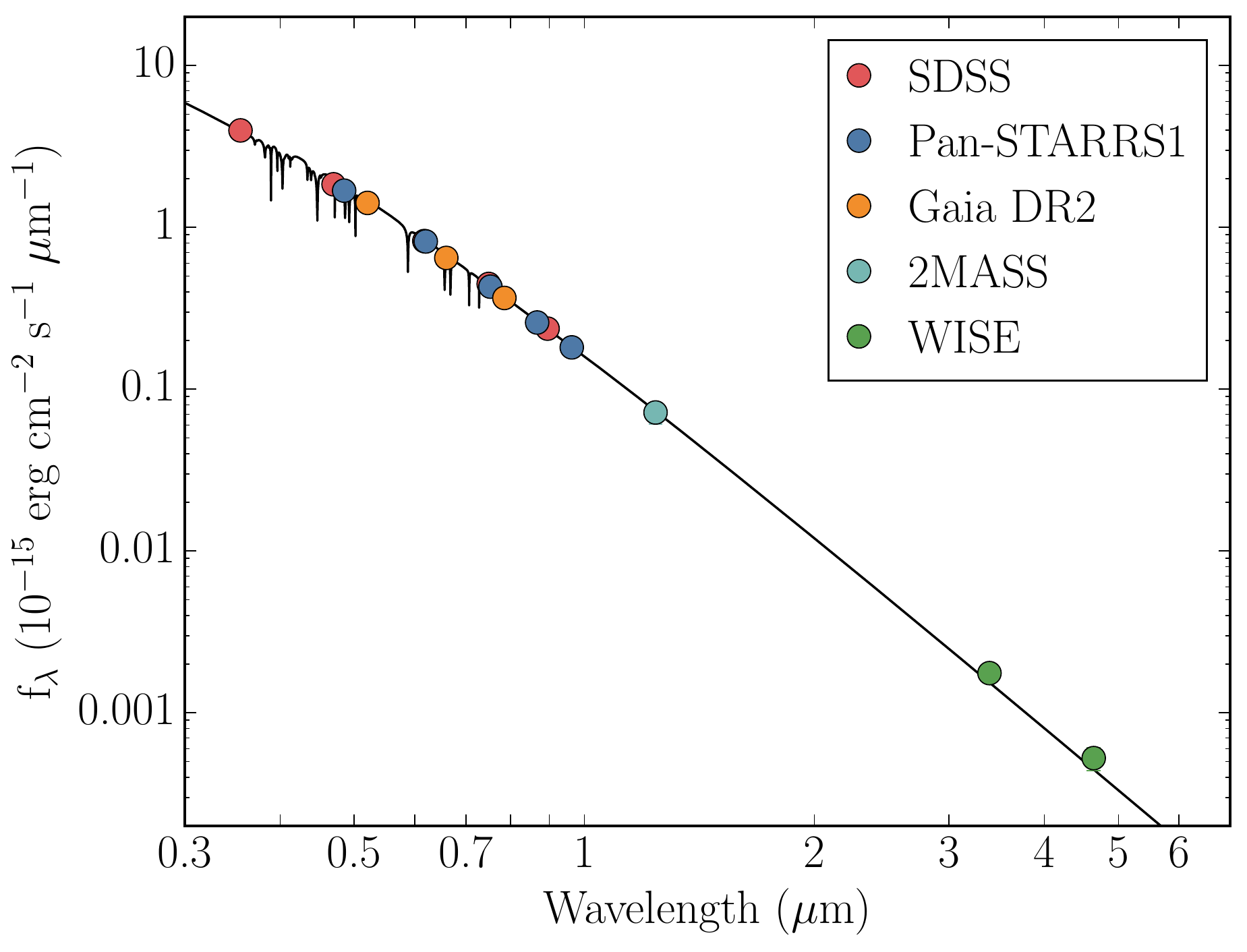}
    \caption{Spectral energy distribution of \target\ with the photometric best-fit white dwarf model (solid line) superimposed ($T_\mathrm{eff}= 16\,913$ K, $\log g= 8.28$~dex and $\htohe = -3.68$~dex, see Section~\ref{sec:justphot}), illustrating the absence of an infrared excess. The synthetic spectrum has been scaled with the radius of the white dwarf ($\Rwd=0.0109\,\Rsun$) and the \textit{Gaia} DR2 parallax of the source, and reddened using $E(B-V)=0.01$. The optical photometry points shown are from Table~\ref{tab:phot_points}. The infrared data points come from 2MASS  ($J=16.585 \pm 0.151$\,Vega mag, \citealt{skrutskieetal06-1}) and \textit{WISE} ($W_{1} = 16.629 \pm 0.059$ and $W_{2} = 16.629 \pm 0.163$\,Vega mag, \citealt{cutrietal14-1}).}
    \label{fig:SED}
\end{figure}

\target\ is a metal-polluted, helium-dominated white dwarf that can be in any of the three accretion phases mentioned above. Inspection of the spectral energy distribution of \target\ reveals no infrared excess (Fig.~\ref{fig:SED}), that can be interpreted as a lack of a debris disc, i.e. a reservoir of material that can be accreted. While this could be taken as an argument for the system being in the decreasing state, there is growing evidence that some discs may be too faint to be detected (\citealt{bergfors14, wilsonetal19}; see also the discussion by \citealt{bonsor17}). 

We have computed $M_\mathrm{CVZ}$, $v_\mathrm{z}$ and $\tau_\mathrm{z}$\footnote{$\tau_\mathrm{z}$ depends on $M_\mathrm{CVZ}$, the density and depth of the CVZ, and $v_\mathrm{z}$.} for each detected metal following \citet{koester09-1}. We used the updated input physics for envelopes and diffusion of \cite{koester20} and the absolute lower limit on the total mass accreted (see Table\,\ref{tab:metalabs}) as the sum of the metal masses in the CVZ obtained with: 

\begin{equation}
M_{\mathrm{z}} = M_{\mathrm{CVZ}}\ Z_{\mathrm{ph,m}}(t)~.
\label{eq:m_z}
\end{equation}

\noindent
In the following, we discuss each of the possible accretion states and derive the corresponding parent body abundances, that we compare with those of planetary bodies in the inner Solar System and several DAZ and DB[A]Z white dwarfs with reliably inferred parent body abundances (i.e. which are unequivocally in the steady state).

\begin{table}
\centering
\caption{Absolute lower limits on the metal masses contained in the CVZ ($M_\mathrm{z}$, $M_\mathrm{z,ds}$) and accretion rates $\dot M_\mathrm{z}$, assuming that \target\ is in the increasing, steady, or decreasing state (is, ss, and ds, respectively). The $M_\mathrm{z}$ values are computed using Eq.~\ref{eq:m_z} with the measured $Z_{\mathrm{ph,m}}$ (Table~\ref{tab:metalabs}) and are \textit{independent} on the accretion state. Accretion rates (which reflect the parent body abundances) are computed from Eq.~\ref{eq:bus} for the increasing state assuming that accretion started 1$\times \tau_{\mathrm{Si}}$ ago, and from Eq.~\ref{eq:ss} for the steady state. If \target\ is in the decreasing state, there will be no ongoing mass transfer. Instead, we computed $\dot M_\mathrm{z,ds}$ corresponding to the steady state \textit{before} accretion switched off. For that purpose, we corrected the measured abundances with $\tau_{\mathrm{z}}$, assuming that accretion stopped 3$\times \tau_{\mathrm{Si}}$ ago (Eq.~\ref{eq:ds}), and computed $\dot M_\mathrm{z,ds}$ using Eq.~\ref{eq:ss} and $M_\mathrm{z,ds}$ with Eq.~\ref{eq:m_z}. }
\begin{tabular}{cccccc}
\hline
Element & $M_{\mathrm{z}}$   &   & $\dot{M}_{\mathrm{z}}$  &  & $M_{\mathrm{z,ds}}$  \\
 & ($10^{20}$ g)  &   & ($10^{8}$ g s$^{-1}$) &  & ($10^{20}$ g) \\
 \noalign{\smallskip}
 & & is & ss & ds & \\
\hline
\noalign{\smallskip}
H$^{\dag}$  & 133  &   -     &   -  &    -     & - \\
O           & 243  &  128 & 63.4 & 487  & 1874\\
Na          & < 4.3 & < 2.5 & < 1.5  & < 23 & < 67\\
Mg          & 102  &  59  & 36  & 554    & 1586\\
Al          & 8  &  4.8 & 3.1 & 64     & 165\\ 
Si          & 85   &  51  & 33  & 654    & 1712\\    
Ca          & 17   &  11  & 8.3  & 388    & 791\\ 
Ti          & 0.5  &  0.4 & 0.3 & 30     & 50\\
Cr          & 2.0  &  1.5 & 1.2 & 171    & 273\\   
Mn          & 0.7  &  0.6 & 0.5 & 86    & 131\\   
Fe          & 98   &  78  & 64  & 10913  & 16683\\ 
Ni          & 4.1  &  3.3 & 2.7 & 513    & 769\\  
\hline
Total & 698 & 340 & 214 & 13883 & 24101\\
\hline
\end{tabular}
\parbox{0.9\columnwidth}{$^{\dag}$No mass accretion rate computed since hydrogen never diffuses out of the photosphere.}
\label{tab:mass_rates}
\end{table}

\subsubsection{\textit{Increasing} state}
We assume that \target\ had an atmosphere composed solely of helium and hydrogen before the onset of accretion. Once mass transfer starts, the accreted material is rapidly and homogeneously mixed throughout the CVZ, but diffuses out of it on a much longer time-scale. Since the first term on the right side of Eq.~\ref{eq:mass_ab} vanishes under this assumption ($Z_\mathrm{ph,m}(0)=0$), the time-dependent mass abundance of each metal in the CVZ during the increasing state is:

\begin{equation}
Z_{\mathrm{ph,m}} (t) =  \frac{\tau_{\mathrm{z}}\dot{M_{\mathrm{z}}}}{M_{\mathrm{CVZ}}}\left[1-e^{-t/\tau_{\mathrm{z}}}\right]~.
\label{eq:bus}
\end{equation}

\noindent
At the very beginning, the abundances within the whole CVZ, and hence also those measured from the photospheric spectrum, will be identical to those of the parent body. As time goes on, the heavier elements start to diffuse out faster compared to the lighter ones, and their relative abundances in the photosphere drop until accretion/diffusion equilibrium is reached after about five diffusion time-scales. At this point, the photospheric abundances of the elements that diffuse out faster become \textit{lower} than in the parent body. 

\begin{figure*}
\centering
\includegraphics[width=2.09\columnwidth]{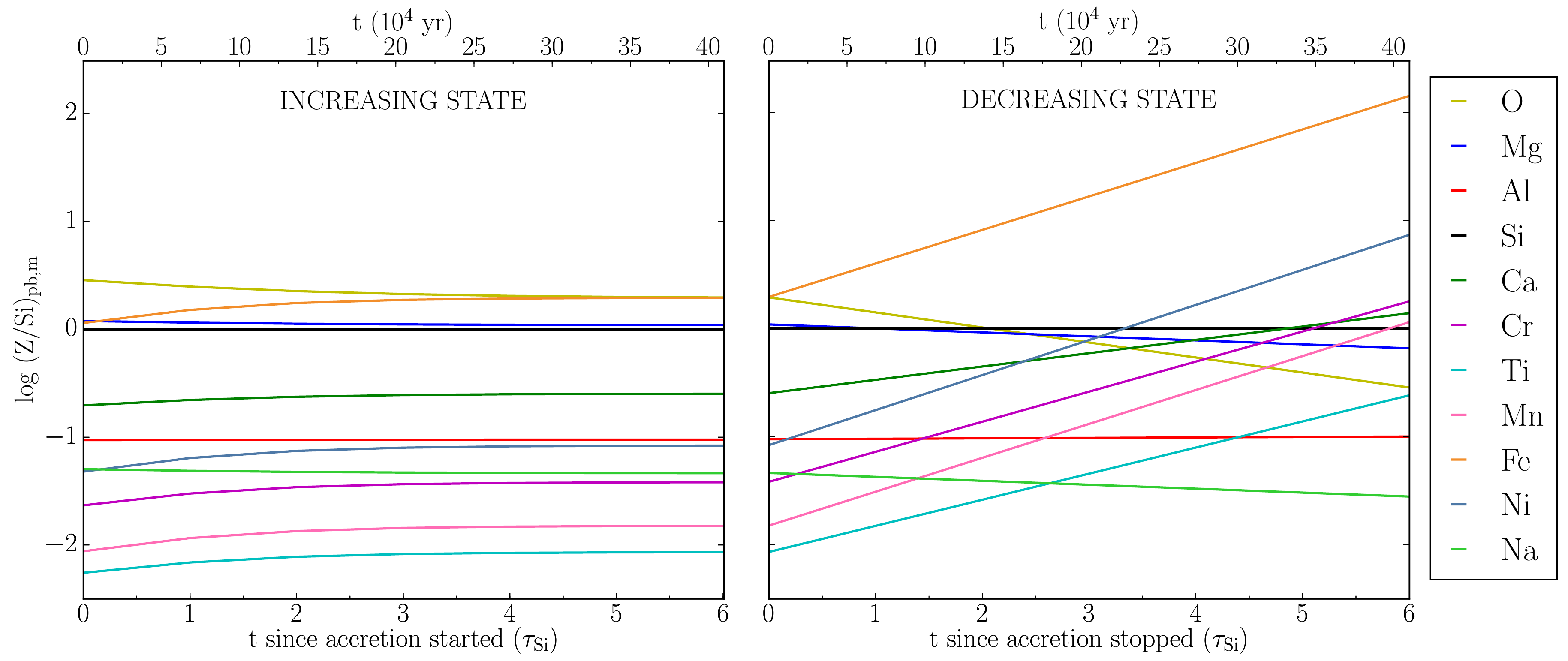}
\caption{Deviation in the inferred parent body mass abundances given the uncertainty in the \target\ accretion state. We assumed a simple scenario in which mass transfer switches on, remains constant for some time and then switches off, so the white dwarf can be found in any of three different states: the increasing, the steady, and the decreasing state. The elapsed time since accretion started or stopped (left and right panel, respectively) is measured in silicon diffusion time-scales. Note that the parent body abundances converge between the two panels, where the accretion state of the source is the steady state (see Section~\ref{sec:PB_abs} for details). Left panel: during the increasing state, material is accreted on to the white dwarf surface and the  photospheric abundances increase linearly with time and are overall close to those of the parent body. Right panel: during the decreasing state, the photospheric abundances decrease exponentially and proportional to the individual sinking time of each metal, resulting in rapidly increasing changes in the inferred parent body abundances as a function of the time since accretion stopped.}
\label{fig:bus_ds}
\end{figure*}

\begin{figure*}
\centering
\includegraphics[width=1.7\columnwidth]{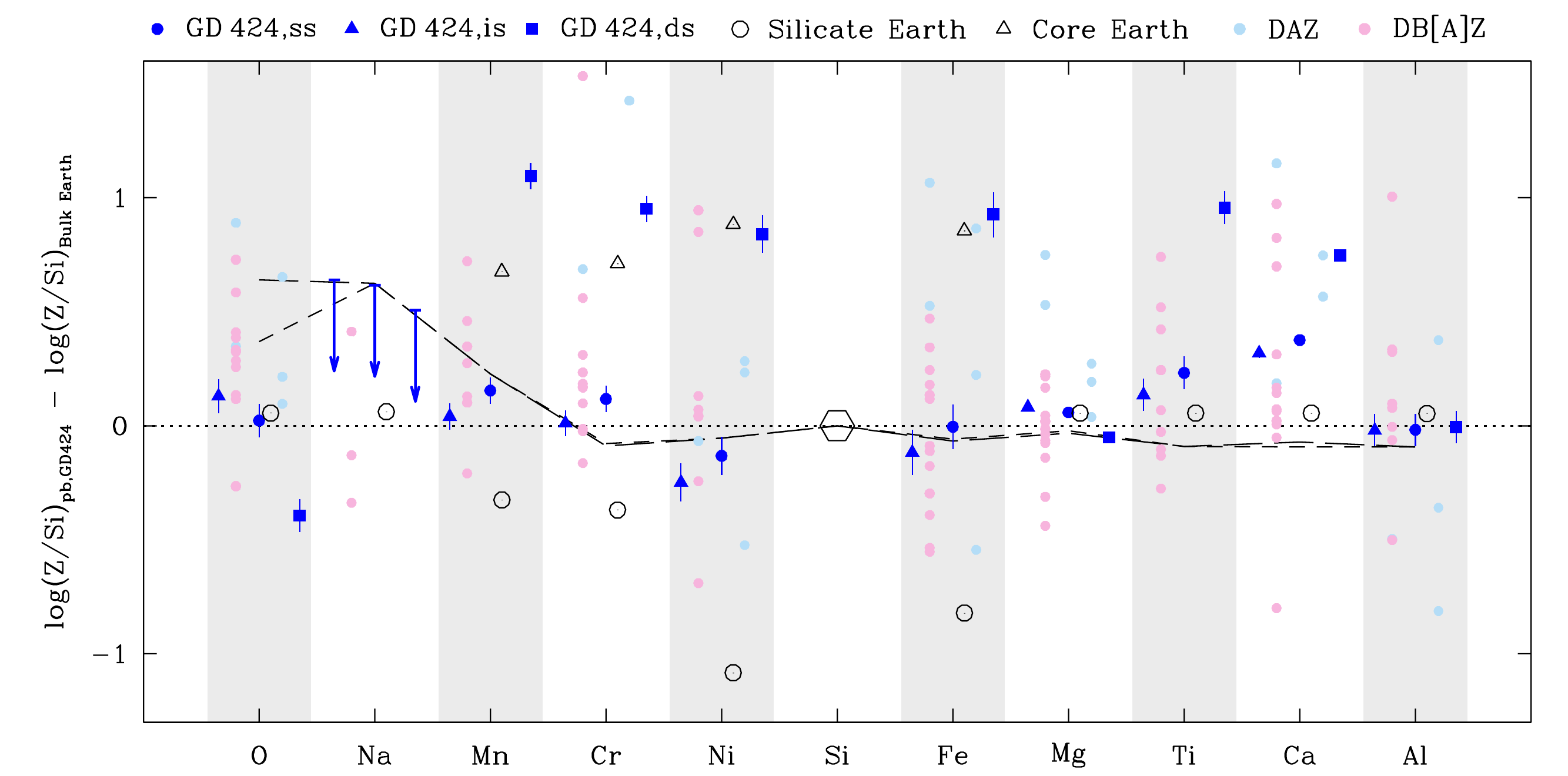}
    \caption{Abundance ratios relative to silicon (the large open hexagon) of the planetary body accreted by \target\ for three different assumptions (is\,=\,increasing state, accretion started $1\times\tau_\mathrm{Si}$ ago; ss\,=\,steady state; ds\,=\,decreasing state, accretion stopped $3\times\tau_\mathrm{Si}$ ago). The abundance ratios are normalized by their corresponding bulk-Earth values \citep{mcdonough00-1} and the elements are sorted by increasing condensation temperature \citep{lodders03-1}.  The long and short dashed lines indicate Solar and CI chondrite abundances, respectively. For comparison, the abundance ratios measured in five DAZ white dwarfs (light blue symbols, from \citealt{gaensicke12} and \citealt{melisetal17-1}), and 17 DB[A]Z white dwarfs (rosa symbols, from \citealt{juraetal12-1, dufouretal12-1, xuetal13-1, farihietal13-2, wilsonetal15-1, raddietal15-1, farihietal16-1, xuetal17-1, xuetal19-1,  fortin-archambaultetal20-1, hoskinetal20-1}) are displayed. The DAZ stars have extremely short diffusion time-scales, and are therefore in the steady state. For the DB[A]Z, the published abundances assuming steady state accretion are shown. If \target\ were actually in the decreasing state, the abundance ratios of the accreted body would radically differ from any object in the Solar System, with large mass fractions of both core elements (Fe, Ni) and crust elements (Ca, Ti).}
    \label{fig:abund_condensation}
\end{figure*}

\begin{figure*}
\centering
 \includegraphics[width=2.0\columnwidth]{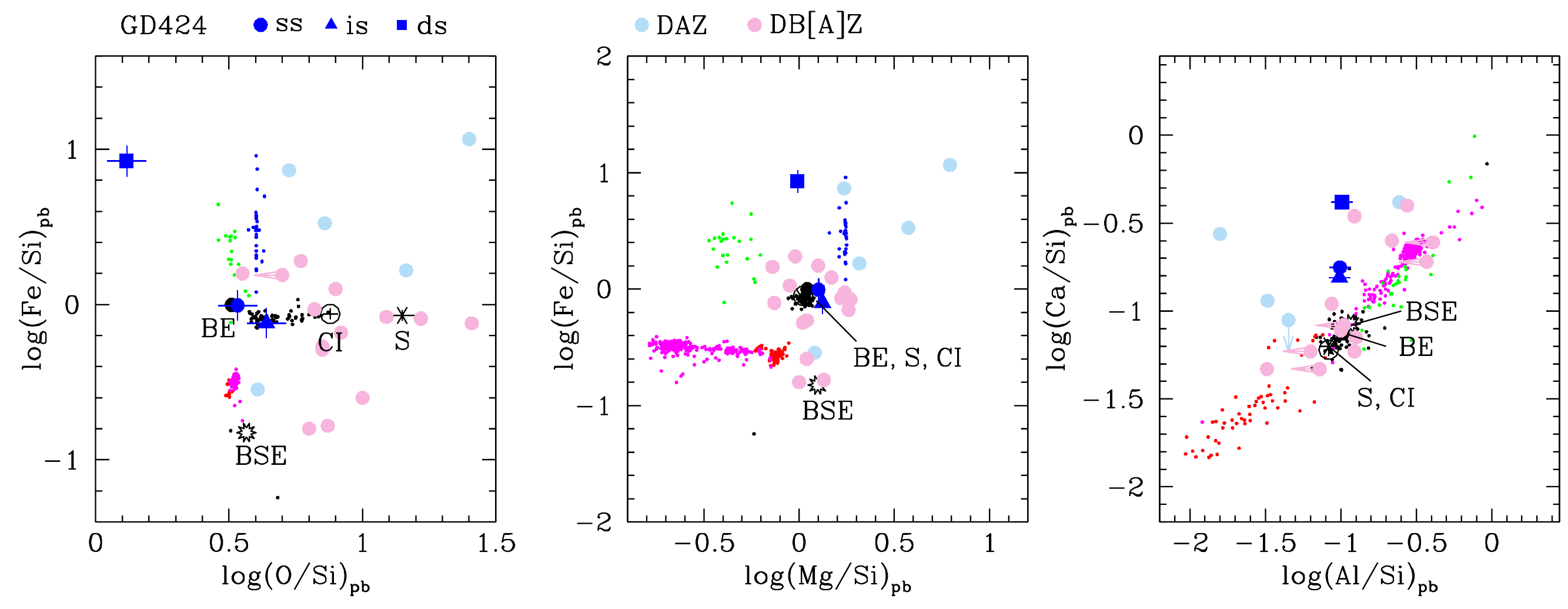}
    \caption{Abundance ratios relative to silicon of the planetesimal accreted by \target\ for three different assumptions (is\,=\,increasing state, accretion started $1\times\tau_\mathrm{Si}$ ago; ss\,=\,steady state; ds\,=\,decreasing state, accretion stopped $3\times\tau_\mathrm{Si}$ ago). Shown for comparison are the abundance ratios of the same DAZ and DB[A]Z as in Fig.\,\ref{fig:abund_condensation}, those of the bulk Earth and bulk silicate Earth (BE and BSE; \citealt{mcdonough00-1}), of the Sun and CI chondrites (S and CI; \citealt{lodders03-1}), and of different types of meteorites (grey\,=\,carbonaceous Chondrites, green\,=\,Mesosiderites, blue\,=\,Pallasites, red\,=\,Diogenites,
  orange\,=\,Howardites, magenta\,=\,Eucrites;
  \citealt{nittleretal04-1}).} 
    \label{fig:abund_ratios}
\end{figure*}

Taking into account the metal sinking times derived for \target, we estimated the parent body mass abundances $Z_{\mathrm{pb,m}}$ (which are mirrored in $\dot{M_{\mathrm{z}}}$) in terms of the time since accretion started (Fig.\,\ref{fig:bus_ds}, left panel). As the system evolves, the abundances of titanium, chromium, manganese, iron and nickel increase by $\approx0.2$\,dex relative to silicon, while sodium, magnesium, aluminium and calcium remain almost equal (they have sinking times similar to silicon). Oxygen is the only element whose abundance decreases by $\approx0.2$\,dex. In the light of this outcome, for the increasing state the uncertainty in the time since accretion started does not result in a big difference on the inferred parent body abundances. In Figs.\,\ref{fig:abund_condensation} and \ref{fig:abund_ratios} we compare the metal abundances of the planetesimal accreted by \target\ (assuming that accretion started one $\tau_{\mathrm{Si}}$ ago; blue triangles) with inner Solar System bodies, CI chondrites, and the debris abundances measured for a number of DAZ and DB[A]Z white dwarfs. We find that the abundances of the parent body resemble CI chondrites, with the only noticeable exception of an overabundance of calcium.



\subsubsection{Steady state}
After about five diffusion time-scales of continued mass transfer, the system reaches an accretion/diffusion equilibrium (Fig.~\ref{fig:bus_ds}, left panel). For $t\gg\tau_\mathrm{z}$, Eq.~\ref{eq:mass_ab} becomes:

\begin{equation}
Z_{\mathrm{ph,m}}(t) = \frac{\tau_{\mathrm{z}}\dot{M_{\mathrm{z}}}}{M_{\mathrm{CVZ}}}~.
\label{eq:ss}
\end{equation}

\noindent
We then inferred the relative parent body abundances assuming a steady state for \target. They are at most within 0.2\,dex of those obtained assuming an increasing state (blue circles in Figs.\,\ref{fig:abund_condensation} and \ref{fig:abund_ratios}), and are consistent with CI chondrites and the bulk Earth. Again, the only outlier is a noticeable calcium overabundance.

\subsubsection{\textit{Decreasing} state}

Once accretion stops, the metals with shorter diffusion times are depleted from the CVZ, and hence from the photosphere, faster. In this state, the second term of Eq.\,\ref{eq:mass_ab} vanishes, and the photospheric mass abundances can be computed as:

\begin{equation}
Z_{\mathrm{ph,m}} (t) = Z_{\mathrm{ph,m}}(0)\,e^{-t/\tau_{\mathrm{z}}}~.
\label{eq:ds}
\end{equation}

\noindent For a given elapsed time, and with the photospheric abundances that we measure today in the spectra of \target\ ($Z_{\mathrm{ph,m}} (t)$, see Table~\ref{tab:metalabs}), we can infer the mass abundances in the steady state before accretion stopped ($Z_{\mathrm{ph,m}}(0)$ in this case), and hence derive the parent body abundances calculating $\dot{M_{\mathrm{z}}}$ through Eq.\,\ref{eq:ss}.

We have computed the relative parent body abundances as a function of time since accretion stopped (Fig.\,\ref{fig:bus_ds}, right panel). In this state, the uncertainty in the precise moment when accretion stopped translates into very diverse parent body compositions. We find that the larger the disparity between sinking times, the greater the difference in relative abundances compared to the steady state values as time goes on. The diffusion times for magnesium, aluminium and sodium are of the order of $\tau_{\mathrm{Si}}$ and, thus, their relative abundances are not substantially affected by the time elapsed since accretion stopped. However, for the rest of the metals there is a $\sim$1-dex abundance change for a $3\times \tau_{\mathrm{Si}}$ difference in diffusion time. This results in drastic over-abundances of both core elements (iron and nickel) and crust elements (calcium and titanium; squares in Figs.\,\ref{fig:abund_condensation} and \ref{fig:abund_ratios}).


\subsubsection{Likely accretion state and parent body properties}

The uncertainty in the time elapsed since accretion started/stopped results in different parent body compositions. For the increasing and steady states, the inferred parent body abundances are very similar and resemble those of CI chondrites and the bulk Earth (Figs.\,\ref{fig:abund_condensation} and \ref{fig:abund_ratios}). The only exception is the overabundance of calcium, which has been detected in a number of other debris-accreting white dwarfs, and tentatively interpreted as signature of heating of the parent bodies \citep{bonsoretal20-1}. However, this is not the case for the decreasing state: as long as minerals form in the same way as in the Solar System \citep[which appears so far to be the case, ][]{doyleetal19-1}, there is no plausible composition of the parent body that could explain the measured photospheric abundances. Based on the inferred parent body abundances for the three states, we argue that \target\ is more likely undergoing accretion, and hence is either in the increasing or the steady state. It is also possible that accretion stopped recently ($t \lesssim 1\times\tau_{\mathrm{Si}} = 8.3\times 10^{4}$\,yr) and the photospheric abundances are still close to those in the steady state. Besides, in Section~\ref{sec:oxygen} we argue that the oxygen budget also points towards \target\ being in the increasing or steady state since it exhibits a huge deficit (up to $870$\,per cent) for the decreasing state, which is inconsistent with the rest of chemical abundances derived for \target.

In Table~\ref{tab:mass_rates} we summarise the accretion properties of \target. The absolute lower limit on the total mass accreted, $7\times10^{22}$\,g, is computed from the measured $Z_{\mathrm{ph,m}}$ (Table~\ref{tab:metalabs}) and is \textit{independent} of the accretion state and comparable to the Solar System asteroid 10~Hygiea \citep{mouretetal07-1}. The total accretion rate is at least $\simeq2\times10^{10}\,\mathrm{g\,s^{-1}}$ (in the steady state, a factor $\approx 1.5$ to 65 higher than in the increasing and decreasing states, respectively). This places the accretion rate of \target\ among the highest observed \citep[see fig.\,5 in][]{bergfors14}.





\subsection{Oxygen balance}
\label{sec:oxygen}
The abundance studies of white dwarfs accreting planetary debris \citep{zuckerman07, gaensicke12, swanetal19-2} have provided conclusive evidence \citep[with very few exceptions, ][]{xuetal17-1} that  the parent bodies are rocky, broadly resembling objects in the inner Solar System \citep{jura+young14-1}. A key aspect in the formation of rocky bodies and their subsequent evolution is the oxygen fugacity of their birth environment. \citet{doyleetal19-1} concluded that the extra-Solar planetesimals studied so far appear to have very similar geochemical properties to the rocky planets and asteroids in the Solar System. 

Given that we detect the major rock-forming elements in \target, and assuming the same mineralogy as in the Solar System, it is hence possible to probe for the water content of the accreted planetesimal. We thereby computed the oxygen budget of the parent body disrupted by \target\ following \citeauthor{klein10} (\citeyear{klein10}, \citeyear{klein11}), i.e. we account for oxygen carried into the white dwarf atmosphere in the form of MgO, Al$_2$O$_3$, SiO$_2$, CaO, TiO$_2$, Cr$_2$O$_3$, MnO, FeO, and NiO. 

Any leftover oxygen would then require an additional carrier. Given the existing detections and upper limits on the volatile content of exo-planetesimals accreted by white dwarfs \citep{gaensicke12, xuetal13-1, xuetal14-1}, water, either in the form of ice or hydrated minerals, is the only plausible explanation for the excess oxygen detected in a small number of systems \citep{farihietal13-2, raddietal15-1}. The main uncertainty in this approach is the fraction of metallic iron and nickel, where a mixture of Fe and FeO, and Ni and NiO, are most likely \citep{doyleetal19-1}. 

Hence, using the parent body abundances derived for the increasing, steady, and decreasing states (Table~\ref{tab:mass_rates}), we computed the oxygen budget for the two extreme scenarios. In the first case, if all the iron and nickel has a metallic origin, we found a 16\,per cent oxygen excess for the increasing state, while the steady and decreasing state showed a 7 and 196\,per cent oxygen deficit, respectively. In the second case, assuming all the iron and nickel in the CVZ of \target\ were accreted in the form of minerals, the oxygen budget translates into a 2, 37 and 870\,per cent oxygen deficit for the increasing, steady and decreasing state, respectively. For context, under the assumption of an entirely mineral origin for Fe and Ni, the bulk Earth has an oxygen deficit of 36\,per cent \citep{mcdonough00-1}, corresponding roughly to the core mass of the Earth.

Most likely, reality lies in between these two extreme scenarios, where iron and nickel were partly acquired from minerals and also as metals. Given the numbers for the oxygen budget above and the close resemblance of the overall abundances of \target\ to those of the bulk Earth (Figs.\,\ref{fig:abund_condensation} and \ref{fig:abund_ratios}), we can rule out that the system has spent a significant amount of time in the decreasing state.Hence, \target\ is most likely accreting dry, rocky debris in either the increasing or steady state.

\subsection{Trace hydrogen}
A large fraction of all DB white dwarfs display trace amounts of photospheric hydrogen, with both the abundance and total mass of hydrogen in the CVZ broadly showing an increasing trend with growing cooling age \citep{vossetal07-1, bergeronetal11-1}. Hypotheses put forward to explain the observations of these DBA white dwarfs include the mixing of residual hydrogen left over from the evolution of the white dwarf progenitor \citep{koester+kepler15-1}, accretion from the interstellar medium \citep{alcock+illarionov80-1}, accretion of water-rich planetary debris \citep{gentile17}, and hydrogen dredge-up from deeper envelope layers \citep{rollandetal20-1}. Interstellar accretion has been ruled out on the basis of various reasons \citep{friedrichetal04-1, farihietal10-2}, and too little detailed work is so far available on the dredge-up scenario, so that we do not discuss those two possibilities.

Both the hydrogen abundance of \target, $\log(\mathrm{H/He})=-3.65$\,dex, and the total mass of hydrogen contained in the CVZ, $M_{\mathrm{H}}=1.33 \times 10^{22}$\,g are, for its temperature, among the highest values found for the known DBA white dwarfs (see fig.\,5 in \citealt{rolland18} and fig.\,6 in \citealt{gentile17}, respectively). Such a large amount of hydrogen is incompatible with both the convective dilution of residual hydrogen (see figs.\,14 and 7 of \citealt{rolland18} and \citealt{rollandetal20-1}, respectively) and accretion at a quasi-steady rate (see figs.\,15 and 8 of \citealt{rolland18} and \citealt{rollandetal20-1}, respectively): \target\ would have evolved into a DA white dwarf.

As noted by \citet{rolland18}, the only possible way to explain the large amount of trace hydrogen in DBA white dwarfs such as \target\ is that accretion only occurred \textit{after} the white dwarf has cooled enough to develop a CVZ that is sufficiently deep to maintain the hydrogen mixed. Gauging from the reflection points in fig.\,15 in \citet{rolland18}, we estimate that the hydrogen present in the CVZ of \target\ today must have been accreted within the last $50-100$\,Myr. \target\ follows the correlation found by \cite{gentile17} that DB white dwarfs with large amounts of trace hydrogen are also contaminated by metals, corroborating that accretion of planetary debris is indeed at least contributing to the hydrogen detected in their CVZ\footnote{Water within icy planetesimals is likely to survive  the evolution of the white dwarf progenitor through the giant branches \citep{malamud+perets16-1, malamud+perets17-1, malamud+perets17-2}.}.

However, from the discussion in the previous subsections, it is apparent that \target\ is currently accreting the debris of a dry parent body, which does not contribute to the $M_\mathrm{H}$ in the envelope. Thus, a planetary origin of $M_\mathrm{H}$ in \target\ requires previous accretion of water-rich material. 

Assuming that all the hydrogen in the CVZ of \target\ was accreted in the form of water implies a total mass of water of $M_\mathrm{H_2O}\simeq1.19\times10^{23}$\,g, and hence an associated oxygen mass of $M_\mathrm{O}\simeq1.06\times10^{23}$\,g. Whereas this is an order of magnitude larger than the amount of oxygen currently held in the CVZ, because of the short diffusion time scale of oxygen ($1.22\times10^{5}$\,yr), a large, water-rich body might have been accreted as little as $\simeq1$\,Myr ago without leaving a noticeable trace of the episode that delivered the hydrogen to \target. So far, only a handful of metal-polluted white dwarfs are known to be under ongoing accretion of water-rich planetesimals \citep{farihietal13-2, raddietal15-1}, and hence it is more likely that part of the $M_\mathrm{H}$ was acquired in just one earlier accretion event of a massive planetesimal rather than in multiple episodes\footnote{A speculative scenario in which $M_\mathrm{H}$ was delivered by multiple much smaller planetesimals  could invoke volatile-rich Kuiper-Belt-like objects \citep{caiazzo+heyl17-1}, possibly perturbed by an unseen wide companion \citep{stephanetal17-1}. However, considering that so far only one white dwarf is known to accrete a volatile-rich planetesimal \citep{xuetal17-1}, and dynamical arguments \citep{veras14}, such a scenario appears unlikely.}.

\section{Conclusions}
\label{sec:conclusions}
We presented the discovery and chemical abundances analysis of \target, a metal-polluted DBA white dwarf with one of the largest amounts of trace hydrogen measured so far among white dwarfs with similar temperatures: $M_{\mathrm{H}}=1.33 \times 10^{22}$\,g.

To determine the photospheric parameters of \target, we used a hybrid method to simultaneously fit a grid of synthetic spectra, survey photometry, and the \textit{Gaia} DR2 parallax to our WHT optical spectrum. This analysis combines the spectroscopic sensitivity to the chemical composition, \htohe, with that to \Teff\ and \logg\ of the photometry. The photospheric parameters are $T_\mathrm{eff}=16\,560 \pm 75$\,K, $\log g=8.25 \pm 0.02$\,dex, and $\htohe=-3.65 \pm 0.03$\,dex. These yield a white dwarf mass, radius, and cooling age of $\Mwd=0.77\pm0.01\,\Msun$, $\Rwd=0.0109\pm0.0001\,\Rsun$, and $\tau_\mathrm{cool}=215\pm10$
~Myr, respectively.

We also obtained high resolution spectroscopy with the Keck telescope. We identified a total of 11 metals in the Keck and WHT spectra of \target: oxygen, sodium, manganese, chromium, nickel, silicon, iron, magnesium, titanium, calcium and aluminium, whose presence we attribute to accretion of a planetary body on to the white dwarf. 

We measured the photospheric metal abundances and used them to estimate the parent body composition. Both the inferred parent body abundances and the oxygen balance suggest that the metals were most likely acquired in a recent or ongoing accretion event, and thus the system is in the increasing or the steady state. The estimated composition of the parent body is consistent with both CI chondrites and the bulk Earth. Hence, future observations to measure the abundance of volatile elements are encouraged to differentiate between these two alternatives. 

The composition of the parent body did not reveal an oxygen excess. This suggests that the large amount of trace hydrogen is probably the result of the earlier accretion of water-rich planetesimals.


\section*{Acknowledgements}

PI acknowledges financial support from the Spanish Ministry of Economy and Competitiveness (MINECO) under the 2015 Severo Ochoa Programme MINECO SEV--2015--0548. OT was supported by a Leverhulme Trust Research Project Grant. BTG was supported by the UK STFC grant ST/T000406/1 and by a Leverhulme Research Fellowship. PR-G acknowledges support from the State Research Agency (AEI) of the Spanish Ministry of Science, Innovation and Universities (MCIU) and the European Regional Development Fund (FEDER) under grant AYA2017--83383--P. This paper is based on observations made in the Observatorios de Canarias del IAC with the William Herschel Telescope operated on the island of La Palma by the Isaac Newton Group of Telescopes in the Observatorio del Roque de los Muchachos. Some of the data presented herein were obtained at the W.~M. Keck Observatory from telescope time allocated to the National Aeronautics and Space Administration (NASA; program 60/2019B\_N094, PI Redfield). We acknowledge the use of Tom Marsh's {\sc pamela} and {\sc molly}. This research has made use of the SVO Filter Profile Service (\url{http://svo2.cab.inta-csic.es/theory/fps/}) supported from the Spanish MINECO through grant AYA2017--84089--P. 

\section*{Data Availability Statement}

The data underlying this article will be shared on reasonable request to the corresponding author.




\bibliographystyle{mnras}
\bibliography{gd424_paper}







\bsp	
\label{lastpage}
\end{document}